\documentstyle[11pt,aaspp4]{article}

\def\chisq{\chi^{2}}
\def\cm3{\rm cm^{-3}}

\def\h2{{\rm H}_{2}}
\def\trot{{\rm T}_{rot}}
\def\tvib{{\rm T}_{vib}}

\begin{document}

\begin{center}
Submitted to {\it The Astrophysical Journal}
\end{center}

\vskip 24pt

\title{Near-Infrared Spectroscopy of Molecular Hydrogen Emission in Four 
Reflection Nebulae: NGC~1333, NGC~2023, NGC~2068, and NGC~7023}

\vskip 24pt

\author{Paul Martini, K. Sellgren, \& D.L. DePoy}

\vskip 12pt

\affil{Department of Astronomy, 174 W. 18th Ave., Ohio State University, \\
Columbus, OH 43210 \\
martini,sellgren,depoy@astronomy.ohio-state.edu}

\vskip 12pt

\centerline
{\bf Abstract}

We present near-infrared spectroscopy of fluorescent molecular hydrogen 
($\h2$) emission from 
NGC~1333, NGC~2023, NGC~2068, and NGC~7023 and derive the physical properties
of the molecular material in these reflection nebulae. Our observations of 
NGC~2023 and NGC~7023 and the physical parameters we derive for these nebulae 
are in good agreement with previous studies. Both NGC~1333 and NGC~2068 have 
no previously-published analysis of near-infrared spectra. Our study reveals 
that the rotational-vibrational states of molecular hydrogen in NGC~1333 are 
populated quite differently from NGC~2023 and NGC~7023. We determine that the 
relatively weak UV field illuminating NGC~1333 is the primary cause of the 
difference.  Further, we find that the density of the emitting material in 
NGC~1333 is of much lower density, with $n \sim 10^2 - 10^4\;\cm3$. NGC~2068 
has molecular hydrogen line ratios more similar to those of NGC~7023 and 
NGC~2023.  Our model fits to this nebula show that the bright, $\h2-$emitting 
material may have a density as high as $n \sim 10^5\;\cm3$, similar to what we 
find for NGC~2023 and NGC~7023. 

Our spectra of NGC~2023 and NGC~7023 show significant changes in both the 
near-infrared continuum and $\h2$ intensity along the slit and offsets 
between the peaks of the $\h2$ and continuum emission.  We find that these 
brightness changes may correspond to real changes in the density and 
temperatures of the emitting region, although uncertainties in the total 
column of emitting material along a given line of sight complicates the 
interpretation.  The spatial difference in the peak of the $\h2$ and 
near-infrared continuum peaks in NGC~2023 and NGC~7023 shows that the 
near-infrared continuum is due to a material which can survive closer to 
the star than $\h2$ can.


\clearpage

\section{Introduction}

Molecular hydrogen ($\h2$) is the most abundant molecule in the universe. 
The study of this molecule has applications to a wide range of astrophysical 
questions ranging from the episode of reionization in the early universe
to spectra of ultraluminous IR galaxies to star formation and photodissociation 
regions (PDRs) in our galaxy.  In PDRs, including the reflection nebulae we 
discuss in this paper, observations of $\h2$ line emission can be a valuable
means of extracting physical information about the molecular material in 
areas of young star formation. Models of PDRs (see the recent review
by Hollenbach \& Tielens 1997) include predictions for the relative population 
of the rotational and vibrational levels of $\h2$ as a function of density, 
temperature, and the strength of the ultraviolet ($UV$) radiation field 
(Black \& van Dishoeck 1987; Sternberg \& Dalgarno 1989; Burton, Hollenbach, 
\& Tielens 1990; Draine \& Bertoldi 1996). As $\h2$ is a homonuclear molecule, 
it has no permanent dipole moment and therefore emits primarily via quadrupole 
transitions. The strongest of these transitions primarily occur in the near- 
and mid-infrared regions of the electromagnetic spectrum. The near-infrared 
$K-$band (centered at 2.2 $\mu$m), in particular, contains a number of 
well-separated emission lines and is relatively accessible from ground-based 
observatories. 

Observations of molecular hydrogen line ratios have been used to 
estimate the physical properties of the reflection nebulae NGC~2023 
(Gatley et al. 1987; Hasegawa et al. 1987; Takayanagi et al. 1987; 
Black \& van Dishoeck 1987; Draine \& Bertoldi 1996; Burton et al. 1998)
and NGC~7023 (Martini, Sellgren, \& Hora 1997). This information is valuable
in conjunction with mid-infrared observations of the dust distribution and 
dust spectral features (Gatley et al. 1987; Cesarsky et al. 1996) for 
determining the nature and 
properties of the dust present in these regions. The study of nebulae with a 
range of densities and $UV$ field strengths aids in understanding how physical 
conditions in nebulae affect the dust content, such as by investigating 
correlations between dust properties and the physical properties of the PDRs 
in which it is observed. 

In this paper we present near-infrared spectroscopy of molecular hydrogen
line emission in four reflection nebulae: NGC~1333, NGC~2023, NGC~2068, and
NGC~7023.  The positions we observed in these nebulae were chosen to coincide 
with mid-infrared spectroscopy of these regions with the {\it Infrared Space 
Observatory} (ISO).  These observations include studies of two nebulae, 
NGC~1333 and NGC~2068, in which the $\h2$ emission spectrum has not been 
previously analyzed. For NGC~2023 and NGC~7023, we have observed new slit 
positions and compare these regions with physical parameters derived for other 
regions.  

\section{Observations and Data Reduction}

We obtained the observations discussed in this paper on observing runs in 
September-October 1997 (NGC~7023 and NGC~1333) and January-February 1998
(NGC~2023 and NGC~2068). A log of our observations is presented as Table 1. 
These observations were obtained with the near-infrared imager/spectrometer
OSIRIS (DePoy et al. 1994) on the 1.8m Perkins telescope at Lowell Observatory.
We used the cross-dispersed mode for these observations, which includes a
4\farcs5 $\times 60''$ slit oriented North-South. Slight vignetting by a loose
piece of baffling decreased the effective slit length to $\sim 45''$. This mode 
has a 2-pixel resolution of $R = \lambda/\Delta\lambda = 565$ at 2.2 $\mu$m, 
565 at 1.65 $\mu$m, and 535 at 1.25 $\mu$m. In Figure 1, we present a model 
spectrum of $\h2$ line emission from Draine \& Bertoldi (1996) smoothed to our
resolution. This spectrum shows the rich spectrum of $\h2$ emission expected
from a PDR and also how most of the lines in the $J-$ and $H-$bands are 
blended together in our data. We have marked the strongest lines and blends
on the figure and in Table 2 we list the wavelengths and transitions of the 
lines that contribute to each of these blends. 

We observed these targets in the standard ABBA pattern, where each on-source, 
`B' observation consisted of a 60s exposure and each `A', 60s sky exposure
was obtained on a nearby, blank piece of sky. In addition to our target 
observations, we also observed several atmospheric standards each night within
$\Delta {\rm sec}\;z = 0.1$ of the airmass of our targets. After interpolating
over the hydrogen lines in these stars, they were used 
to remove the wavelength-dependent atmospheric and system transmission. 
We observed these stars by offsetting the telescope north-south 
to move the star along the slit and using the mean, calculated with a 
sigma-rejection algorithm, of the observations to remove the night-sky lines. 
Dome flats, which were the difference of flats obtained with the dome lights on 
and with the lights off, were used to remove variations in the pixel-to-pixel 
sensitivity across the slit. The bright OH night sky lines (Oliva \& Origlia
1992) were used to 
calculate the dispersion solution. We used the A0V atmospheric standards
observed each night to place the counts in the $J-$, $H-$, and $K-$bands 
on a relative flux scale. We calculated the relative flux coefficients by 
fitting these stars with a 9600 K blackbody. This conversion varied little 
from night to night and we estimate the net uncertainty to be 5\% in this 
conversion.  We obtained an absolute calibration for the 1-0 S(1) line 
intensity by placing synthesized apertures on our unpublished broadband and 
narrowband imaging data. 

\section{Data}

\subsection{NGC~1333}

NGC~1333, also known as vdB~17 (van den Bergh 1966), is illuminated by 
BD~$+30\ 549$ (Greenstein 1948; Racine 1968). Its spectral class is B8-9p 
based on photoelectric photometry (Johnson 1960), B8V based
on a spectrum (Racine 1968), or B6 (Harvey, Wilking, \& Joy 1984) based on 
its luminosity. Racine (1968) measured the extinction towards BD~$+30\ 549$ to 
be $A_V = 1.86$ with $R_V = 3.1$. The distance to NGC~1333 is estimated to be 
500~pc (Strom, Grasdalen, \& Strom 1975). We note that our observed position 
in NGC~1333 is different from the better studied and more luminous 
NGC~1333/SVS-3 (e.g. Joblin et al. 1996; Bregman et al. 1993). 

Sellgren (1986) first detected Q-branch $\h2$ emission near 2.4 $\mu$m from 
the reflection nebulosity in this region at a position $20''$ south of 
BD~$+30\ 549$ 
with a $12''$ diameter beam. $\h2$ emission from the 1-0~S(1) and 2-1~S(1) 
lines was detected by Tanaka et al. (1989) at a position $40''$ south of the 
central star with a beam diameter of $19.6''$. Our long slit, oriented 
north-south and centered at $12''$ east $26''$ south of BD~$+30\ 549$, 
overlaps the position observed by Tanaka et al. (1989). 

Our spectrum of NGC~1333 includes detections of a number of lines in the 
near-infrared $K-$band, as well as a few of the more prominent, unblended 
lines in $J$ and $H$. Before summing along the slit, we searched for any 
obvious structure in the $\h2$ line intensities. As 
suggested by our unpublished imaging data, we detected no strong, filamentary 
structure in our slit and we therefore summed the full, useful ($4.5'' \times
45''$) range to produce the spectrum shown in Figure 2. The relative line 
intensities, uncorrected for extinction, are listed in Table 3. Our narrowband
imaging is not sensitive enough to directly measure the 1-0~S(1) line
intensity. Instead, we measured the intensity of the continuum emission and 
used this to calibrate the strength of the 1-0~S(1) line from our spectrum 
to be $1 \pm 0.3 \times 10^{-5}$ erg s$^{-1}$ cm$^{-2}$ sr$^{-1}$. 

\subsection{NGC~2023}

NGC~2023 is one of the brightest reflection nebulae in the sky and was the 
first source in which fluorescent $\h2$ emission was detected (Gatley et al. 
1987; Sellgren 1986). This nebula, also known as vdB~52, is illuminated by
the bright B1.5V star (Racine 1968) HD~37903. NGC~2023 is in the molecular
cloud L1630 and is at a distance of $\sim 400$ pc (Anthony-Twarog 1982). 
Racine (1968) estimated the extinction towards HD~37903 to be $A_V = 1.12$ 
with $R_V = 3.1$.

Gatley et al. (1987) mapped the nebula in the 1-0~S(1) line of $\h2$ and 
obtained low-resolution spectra at positions 80$''$ south and 160$''$ north of 
the HD~37903. These spectra showed a large number of features, including 
the 1-0 Q-branch in the $K-$band and the 6-4 and 5-3 Q-branches in the 
$H-$band. These spectra were also studied by Hasegawa et al. (1987) and 
Takayanagi et al. (1987). Sellgren (1986) detected Q-branch emission from 
positions $60''$ and $80''$ south of HD~37903. More $\h2$ lines have been 
observed in spectra obtained by Burton (1992) and described in more detail by 
Burton et al. (1998). Near-infrared images of the spatial distribution of 
$\h2$ through narrowband filters have been obtained by Field et al. (1994), 
Rouan et al. (1997), and Field et al. (1998). 

We observed NGC~2023 at the position 60$''$ south of HD~37903 observed by
Sellgren (1986). Our long slit also includes the region 80$''$ south observed
by Gatley et al. (1987). Our slit position is near, but to the east of the 
position observed by Burton et al. (1998). The Burton et al. position is 
centered on a peak in the $\h2$ emission located 78$''$ south, 9$''$ west 
of HD~37903 and was observed with a roughly 5$''$ diameter beam. Our spectrum 
extends from approximately 35$''$ to 84$''$ directly south of HD~37903. 

We found that the relative strength of the $\h2$ and continuum emission 
changes significantly along the length of our slit with the region from 
$35''$ to $67''$ south (hereafter the north region) showing strong 
continuum emission and weak $\h2$ lines. The region from $67''$ to $84''$ 
south (hereafter the south region), in contrast, 
shows negligible continuum emission but bright $\h2$ emission. In Figure 3a 
we plot the relative number of counts in the 1-0~S(1) line and the neighboring 
continuum as a function of position along the slit. Because of the significant 
differences in these two regions, we extracted them separately and show the 
north and south regions as Figures 4 and 5, respectively. 
The relative line intensities for both of these regions are listed in Table 3. 
As we do not have narrowband imaging data on NGC~2023, we used $K'$ imaging 
data to measure the 1-0~S(1) line intensity.  We did this by measuring the 
total flux in the $K'$ band and then using our spectra to measure the fraction 
of the $K'$ flux in the 1-0~S(1) line. The absolute 1-0~S(1) line intensity 
for the north region is $8 \pm 3 \times 10^{-5}$ erg s$^{-1}$ cm$^{-2}$ 
sr$^{-1}$, for the south region it is $1.2 \pm 0.4 \times 10^{-4}$ erg 
s$^{-1}$ cm$^{-2}$ sr$^{-1}$. 

\subsection{NGC~2068}

NGC~2068, which is also known as vdB~59, is in the same molecular cloud 
(L1630) as NGC~2023 and is also estimated to be at a distance of approximately 
400 pc (Anthony-Twarog 1982). Strom et al. (1975) classified the central star 
of this nebula, HD~38563-N as a B2III. Strom, Strom, \& Vrba (1976) estimate 
the extinction towards HD~38563-N  to be $A_V = 4$
based on $E(H-K)$. Recent imaging (Martini, DePoy, \& Sellgren 1999) 
shows that HD~38563-N is actually a close ($\sim 2''$ separation) pair of 
stars with comparable $JHK$ brightnesses. A low-resolution $JHK$ spectrum 
($R \sim 750$) of these stars obtained on 9 Sept 1998 shows that both stars 
are consistent with an early B-type classification based on the presence 
of weak hydrogen absorption features. 

Sellgren (1986) unsuccessfully searched for $\h2$ emission from the Q-branch 
at a position $60''$ east, $40''$ south of HD~38563-N. Our unpublished 
narrowband images show no extended $\h2$ emission at this offset position. 
Based on these images we instead selected two positions, centered $72''$ 
and $121''$ east of HD~38563-N as the peaks of the continuum and $\h2$ 
emission, respectively. Throughout this paper we will refer to these positions 
as the ``C-Peak'' and the ``$\h2-$Peak,'' respectively. 

We only detected weak $\h2$ line emission at the C-Peak $72''$ east of 
HD~38563-N, even when the entire slit is averaged together. The spectrum of 
this region is shown in Figure 6. The bright line at 1.64 $\mu$m is probably 
[Fe II] $\lambda 1.644 \mu$m. In principle, the strength of this line can be 
compared to the strength of other [Fe II] lines to derive densities or 
measure the extinction in the emitting region (e.g. Nussbaumer \& Storey 
1988).  However, no other [Fe II] lines were detected in our spectrum and we 
can set no meaningful limits on the density or extinction at this slit 
position.  We note, however, that [Fe II] emission arises in partially or 
fully ionized gas in PDRs. The presence of [Fe II] and $\h2$ in 
this spectrum suggests we are observing at least two different emitting 
regions along the line-of-sight, with ionized gas near the central star and 
$\h2$ emission from molecular material either in front of or behind the 
ionized gas. 

Our slit position centered on the $\h2-$Peak shows much stronger emission. 
Figure 7 is a sum of the slit over the $23''$ region from $0''$ to $23''$ 
north, $121''$ east of the central star. The relative line intensities 
corresponding to this region are listed in Table~3. The southern part of this 
slit position extends beyond the $\h2$ emission filament into a region 
containing no strong $\h2$ or continuum emission. For the $\h2-$Peak, we
measure a 1-0 S(1) line intensity of $3.1 \pm 0.8 \times 10^{-5}$ erg s$^{-1}$ 
cm$^{-2}$ sr$^{-1}$.  We did not clearly detect 1-0~S(1) line emission at the 
continuum region in our narrowband data.  Instead, we used the intensity of 
the continuum region to calibrate the 1-0~S(1) line in our spectrum, as we 
did for NGC~1333. We find the 1-0~S(1) line intensity is 
$7 \pm 3 \times 10^{-6}$ erg s$^{-1}$ cm$^{-2}$ sr$^{-1}$.

\subsection{NGC~7023}

NGC~7023, also known as vdB~139, is a bright reflection nebula illuminated 
by the pre-main-sequence B3e star HD~200775 (Johnson 1960; Racine 1968; 
Witt \& Cottrell 
1980). The distance to NGC~7023 is estimated to be 440 pc (Whitcomb et al. 
1981). We adopt a visual extinction of $A_V = 2.2$ with $R_V = 5$ as 
discussed in Martini et al. (1997). Observation of the $\h2$ Q-branch 
emission from NGC~7023 was first 
attempted, though unsuccessfully, by Sellgren (1986) at a position $30''$ 
west and $20''$ north of HD~200775. Narrowband imaging in the 1-0~S(1) and 
2-1~S(1) lines (Lemaire et al. 1996) later revealed the presence 
of fluorescently excited $\h2$ in the bright filaments seen in this nebula 
at 2.1 $\mu$m by Sellgren, Werner, \& Dinerstein (1992). Near-infrared 
spectroscopy of two of these filaments by Martini et al. (1997) 
showed that the bulk of this emission was due to $UV$ fluorescence in 
filaments with densities of $10^4 - 10^6\;\cm3 $. 

Our observations are centered on a position $27''$ west, $34''$ north of 
HD~200775. This position is close to, but not overlapping, the slit 
position ``P1'' (40$''$ west, 34$''$ north) of Martini et al.  (1997). 
As in the case of NGC~2023, the 
spectrum of NGC~7023 shows significant changes in both continuum and $\h2$ 
line intensity across the slit. The slit is centered on a bright filament, 
where the total ($\h2+$continuum) emission is brightest (hereafter the 
$\h2-$peak region, $32 - 41''$ north). The spectrum of this region is shown in 
Figure 8.  The $\h2$ and continuum emission have noticeably different behavior 
on either side of this peak. The $\h2$ emission lines are still fairly strong 
in the region 
$41 - 56''$ north (hereafter the north region; see Figure 9) of the filament, 
while they are significantly weaker in the region $12 - 32''$ to the south 
(hereafter the south region; not shown) of the filament, closer to HD~200775.  
The relative line intensities for the peak and north regions are listed in 
Table 3. In contrast, the continuum emission is brightest in the south region, 
somewhat less bright at the $\h2-$peak, and quite weak in the north region. 
The behavior of the 1-0~S(1) line and the neighboring continuum are shown in 
Figure 3b.  For the $\h2-$Peak, we measure a 1-0 S(1) line absolute intensity 
of $6.8 \pm 1.7 \times 10^{-5}$ erg s$^{-1}$ cm$^{-2}$ sr$^{-1}$. For the 
north region, the 1-0 S(1) line intensity is $4 \pm 1 \times 10^{-5}$ erg 
s$^{-1}$ cm$^{-2}$ sr$^{-1}$. We note our improved narrowband calibration 
results in significantly lower line intensities for the positions 
``P1'' and ``P2'' reported in Martini et al. (1997). For these positions we 
obtain intensities of $8 \pm 2 \times 10^{-5}$ erg s$^{-1}$ cm$^{-2}$ sr$^{-1}$ 
and $4.6 \pm 1.2 \times 10^{-5}$ erg s$^{-1}$ cm$^{-2}$ sr$^{-1}$, 
respectively; this new flux calibration does not change any of the conclusions 
of Martini et al. (1997).

\section{Results}

In order to derive information about the densities, temperatures, and incident
$UV$ fields in these PDRs, we compared the relative level populations 
of the rotational-vibrational levels of $\h2$ derived from our spectra with 
the model predictions of Draine \& Bertoldi (1996). As these models are 
all characterized by $UV$-field-to-density ratios ($G_0/n_H$) of 0.1 or 0.01, 
we report our best-fitting density models at a given $G_0/n_H = 0.1$ or 
$G_0/n_H = 0.01$ ratio (note that Draine \& Bertoldi 1996 denote the $UV$ 
field strength as $\chi$). $G_0$ is a unit of measure of the $UV$ field 
strength in units of the average interstellar radiation field at $\lambda = 
1000$ \AA\ as defined by Habing (1968). A value of $G_0 = 1$ implies a $UV$ 
field strength of $4 \times 10^{-14}$ erg $\cm3$ (see also Draine \& Bertoldi 
1996). Our modeling procedure is described in detail in 
Martini et al. (1997), although we have modified 
our technique somewhat for this work. Briefly, we compared our
observations with these models by using the $\chisq$ parameter as a measure 
of the goodness of fit between each model and the data. We used all of the 
lines with relatively good ($>2\sigma$) detections from a given slit position 
in the fitting procedure, though we did not include the 2-1~S(4) line, as it
lies on a strong telluric absorption feature, and the 1-0~S(3) line, as it
lies in a region where the atmospheric$+$filter transmission has very 
steep gradient. 

Before performing these model fits, we corrected 
the observed relative line intensities for the effects of interstellar 
extinction. We used the values for the stellar extinction and reddening laws 
quoted above to deredden our measured line intensities following the procedure 
described by Mathis (1990). However, the nebular extinction may be higher or 
lower than the stellar extinction. Martini et al. (1997) found that varying 
the assumed $A_V$ towards NGC~7023 by a factor of 2 had no significant effect 
on their results. We tested the sensitivity of all of our fits to 
the assumed value of $A_V$ by also running fits with $\Delta A_V = \pm 1$; 
we found reddening variations of this magnitude do not affect our results.  
Ideally, the extinction towards these regions should be computed by measuring 
the line strengths of two lines arising from the same upper level and thus 
whose ratios are determined only by their relative transition probabilities 
and statistical weights. Unfortunately we did not detect such a pair of lines 
in any of our objects. However, extinction is not a severe problem for our 
modeling procedure as most of the features we detect lie in the $K-$band and 
so are both close together in wavelength and relatively insensitive to the 
extinction. 

The main differences between the current work 
and the technique described in Martini et al. (1997) are that we performed the
$\chisq$ fit on the line intensity data, rather than on the relative column 
density in each level population, and that we included the intensities of 
several blends of lines in our fitting procedure. In Figure 10a 
we show the variation of one of these blends, H4, as 
a function of density, $UV$ field strength, and temperature. The intensity 
of this blend relative to the 1-0 S(1) line is sensitive to 
density, though increasing the $UV$ field strength or the gas temperature
tends to mimic the effect of increasing density. The behavior
of this blend is representative of the other blends listed in Table 2 because
all of these blends are dominated by transitions from high vibrational levels. 
Their intensity relative to the 1-0 S(1) transition decreases with 
increasing density as the population of the $v = 1$ level is increased by 
collisions. 

We note that Wolniewicz, Simbotin, \& Dalgarno (1998) have 
recently recalculated the spontaneous electronic quadrupole transition 
probabilities for $\h2$. While these values are a significant improvement
over the work of Turner, Kirby-Docken, \& Dalgarno (1977) for transitions with 
probabilities of $10^{-9}$ or less, their values for the higher probability 
transitions, which include the lines we detect, are not appreciably different. 

In addition to the model fits of all of the detected lines and blends with the 
models, we also looked at the ratios of several select lines that have been 
used in the past as diagnostics of PDRs. These lines include the ratio of the 
1-0~S(1) line to the 2-1~S(1) line and the ratio of the 2-1~S(1) line to the 
6-4~Q(1) line. The 1-0~S(1)/2-1~S(1) line ratio is a good diagnostic of the 
density because the rate of collisional deexcitation of the $v=2$ level 
increases 
faster than that of the $v=1$ level. Thus this ratio increases substantially 
when the density is near the critical density of $\h2$, $n_{crit}=10^5\;\cm3$.
The 2-1~S(1)/6-4~Q(1) line ratio compliments the 1-0~S(1)/2-1~S(1) line ratio 
as it is much less sensitive to collisional deexcitation. Draine \& Bertoldi
(1996) show the behavior of these two line ratios as a function of density, 
$UV$ field strength, and gas temperature for their models (see their Figures
15 and 16). We note, however, that the 2-1~S(1)/6-4~Q(1) line ratio is fairly 
sensitive to reddening compared to the 1-0~S(1)/2-1~S(1) ratio. For example, 
adding $A_V = 5$ magnitudes of reddening with an $R_V = 3.1$ reddening law 
(e.g. Mathis 1990) will decrease the 1-0~S(1)/2-1~S(1) ratio by 5\%, but 
increase the 2-1~S(1)/6-4~Q(1) ratio by 30\%. 

In Figure 10b we plot the ratio of the 1-0 S(7) line to the 
intensity of the H4 blend discussed above. As these two features are very 
close in wavelength, their ratio is very insensitive to extinction and the 
reddening law. An error of $A_V = 5$ magnitudes will only increase this 
ratio by $\sim 1\%$. As in the case of the ratio of the H4 blend to the 
1-0 S(1) line, this ratio combines the H4 intensity, dominated by 
transitions from high vibrational levels, with a transition from the 
$v = 1$ level, which is sensitive to increasing density, particularly as 
the density approaches $n_{crit}$. At low densities ($n < 10^4 \cm3$) when 
this ratio is small, this diagnostic is fairly insensitive to variations in 
$UV$ field strength and gas temperature. At higher densities, however, 
increasing $UV$ field strength and temperature mimic the effect of a higher 
density.

Other diagnostics of the physical conditions in PDRs include the rotational 
temperature ($\trot$), vibrational temperature ($\tvib$), and the ratio of 
ortho$-$ to para$-\h2$ ($\gamma$). The rotational (or vibrational) temperature 
is measured from the slope of a line passing through data with the same 
vibrational (or rotational) upper levels on diagrams of the [ln($N_u/g_u$), 
$T_u$]$-$plane, where $N_u$, $g_u$, and $T_u$ are, respectively, the column 
density, statistical weight, and energy (in Kelvin) of the upper level. 
We present population diagrams derived from our spectra showing the 
distribution of ln($N_u/g_u$) vs. $T_u$ in Figures 11 and 12. Differences 
between $\trot$ and $\tvib$ are an indicator of fluorescently excited $\h2$, 
while purely thermally excited $\h2$ results in equal $\trot$ and $\tvib$ 
(Black \& Dalgarno 1976; Black \& van Dishoeck 1987; Takayanagi, Sakimoto, 
\& Onda 1987; Tanaka et al. 1989; Draine \& Bertoldi 1996). In this paper, 
we measure $\trot$ by fitting a line to [ln($N_u/g_u$, $T_u$] 
for the 1-0~S(0) and 1-0~S(2) lines; we chose this pair of lines because they 
are both bright and are even rotational states, which avoids uncertainty
in $\gamma$. We measure $\tvib$ with the 1-0~S(1) and 2-1~S(1) lines. 

$\h2$ is assumed to form on grain surfaces with $\gamma = 3$ 
(e.g. Spitzer \& Zweibel 1974), but after formation this ratio can change 
with time (Dalgarno, Black, \& Weisheit 1973; Flower \& Watt 1984; Tielens \& 
Allamandola 1987; Hasegawa et al. 1987; Tanaka et al. 1989; Burton et al. 
1992; Chrysostomou et al. 1993; Draine \& Bertoldi 1996). In these nebulae, 
we measure $\gamma$ with the $v = 1$, $J = 2 - 4$ levels. Draine \& Bertoldi
(1996) predict $\gamma = 2 \pm 0.2$ for $v=1, J = 2 - 7$ in their model PDRs
and show that $\gamma$ only weakly depends on density and $UV$ field strength. 
Sternberg \& Neufeld (1999) explain how optical depth effects in the $UV$ 
Lyman and Werner bands result in a lower observed $\gamma$ for the excited 
$\h2$ than the $\gamma = 3$ predicted by theory for gas above 200 K. 

\subsection{NGC~1333}

Our model fits to NGC~1333 show that it is best fit by models with 
$n = 10^2 - 10^3\;\cm3$ for $G_0/n_H = 0.1$ and $n = 10^4\;\cm3$ 
for $G_0/n_H = 0.01$. NGC~1333 is illuminated by a B8V star and the $UV$ 
field is expected to be $G_0 = 140$ (Uchida et al. 1999). The results of 
comparing our spectra with model predictions is thus in reasonable agreement 
with the expected $UV$ field. Warin et al. (1996) derive densities of 
$3 \times 10^{3}$ to $5 \times 10^{4}\;\cm3$ in the central regions of NGC~1333 
based on CO observations, which also agree with our results, as does our 
estimate of the intensity of the 1-0 S(1) line. 

The results of using individual line ratios support our model results. 
The 1-0~S(1)/2-1~S(1), 2-1~S(1)/6-4~Q(1), and 1-0~S(7)/H4 ratios (see 
Table 6) all predict densities of $\sim 10^4\;\cm3$. The 2-1~S(1)/6-4~Q(1)
line ratio allows densities of up to $10^5\;\cm3$, but this line ratio
is also the most sensitive to extinction. The extinction estimate towards 
NGC~1333, and for all of the nebulae discussed here, is based on a 
line-of-sight estimate to the central star. Any additional dust screen towards 
the excited molecular gas would increase the extinction and in turn lead to a 
smaller observed 6-4~Q(1) intensity and drive the observed 2-1~S(1)/6-4~Q(1) 
ratio to a higher value. However, as mentioned above, variations in $\Delta A_V 
= \pm 1$ neither changed the best-fitting model parameters nor the quality of 
the fit. 

Our measurements of $\trot$ and $\tvib$ shows that $\tvib$ is significantly 
higher than $\trot$, a further indicator that the $\h2$ emission in NGC~1333 
is fluorescent. We also measure $\gamma = 1.4 \pm 0.4$. Such a low value is 
in agreement with the low density, $UV$ field, and temperature we derived 
above. However, as we note above, Draine \& Bertoldi (1996) found in their 
model PDRs that $\gamma$ is not a strong constraint on density and $UV$ field 
strength.

\subsection{NGC~2023}

The density of NGC~2023 has been previously estimated to be approximately 
$10^5\;\cm3$ with $G_0 \sim 5000$ using a variety of techniques (Burton et al. 
1990; Fuente et al. 1995; Draine \& Bertoldi 1996; Wyrowski et al. 1997; Field 
et al. 1998). Our analysis of the continuum-bright, or north region, 
yields a best-fit density of $n = 10^4 - 10^5\;\cm3$ for $G_0/n_H = 0.1$ and 
$n = 10^5 - 10^6\;\cm3$ for $G_0/n_H = 0.01$. For the $\h2$ bright, or
south region, the best-fitting models predict a density of $n = 10^5\;\cm3$ 
for $G_0/n_H = 0.1$ and $n = 10^6\;\cm3$ for $G_0/n_H = 0.01$. These values 
are also in agreement with our measurements of the 1-0 S(1) line intensities 
at these two slit positions. 

The relative line ratios of both the north and south regions agree well with a 
$n = 10^5\;\cm3$ model and each other, which is consistent with our more 
extensive model fitting. This also suggests that the overall intensity 
difference (Figure 3) between the $\h2-$bright and continuum-bright regions 
do not mark a significant change in the physical properties of the 
PDR. 

Our measurements of $\trot$ and $\tvib$ at both positions in NGC~2023 are 
in good agreement with the measurements of Hasegawa et al. (1987). These 
results are also consistent with the density and temperature ranges 
discussed above as well as the particular models developed by Draine \& 
Bertoldi (1996) for NGC~2023. Our derived value of $\gamma \sim 2$ is 
also consistent with the ortho-to-para ratio found by Hasegawa et al. (1987)
and the model fit of Draine \& Bertoldi (1996). 

\subsection{NGC~2068}

We modeled both the continuum region and $\h2-$bright filament we observed in 
NGC~2068. The continuum region had significantly fainter line emission than
the $\h2$ filament and our models do not strongly constrain the physical 
conditions from this region. Our best-fit models for the continuum region 
constrain the density to be $n = 10^2 - 10^4\;\cm3$ for $G_0/n_H = 0.1$ and 
$n = 10^2 - 10^5\;\cm3$ for $G_0/n_H = 0.01$. As discussed previously, the 
weak $\h2$ observed towards this region could be from relatively lower 
density material along the line of sight, but farther from the central star. 
In the $\h2-$bright region we are able to constrain the density to be in the 
range $n = 10^3 - 10^5\;\cm3$ for $G_0/n_H = 0.1$ and $n = 10^5 - 10^6\;\cm3$ 
for $G_0/n_H = 0.01$.  Our measurement of the line intensity of the 1-0 S(1) 
line suggests $n = 10^4\;\cm3$ for $G_0/n_H = 0.1$ and $n = 10^5\;\cm3$ 
for $G_0/n_H = 0.01$, which is also in agreement with the CS observations of 
Lada, Evans, \& Falgarone (1997). Uchida et al. (1999) estimate $G_0 = 520$ at 
an angular distance of $93''$ from HD~38563-N, corresponding to $G_0 = 870$ at 
the projected distance of the continuum region and $G_0 = 310$ at the projected 
distance of the $\h2-$bright filament. The values are consistent with all but 
the highest density model fits. Our estimate of the 1-0~S(1) line intensity 
for the continuum region suggests the density is $n = 10^{3} - 10^4\;\cm3$. 

We only detected enough lines to measure the 1-0~S(1)/2-1~S(1) ratio in the 
continuum region. This ratio is consistent with a low density 
$n \leq 10^4\;\cm3$ PDR with $T < 1000$ K. At the $\h2$ peak, we measured 
all three ratios and these values imply a density of $n \sim 10^4\;\cm3$. 
Our measurements of $\trot$, $\tvib$, and $\gamma$ are listed in Table~6 and 
are consistent with the model fits described above. Unfortunately, the 
$\h2-$Peak in NGC~2068 is located due west of the central star and we 
observed this nebula with a slit oriented north-south. We are thus not able 
to examine changes in the PDR as a detailed function of distance from the 
central star (and presumably as a function of $UV$ field strength) as we do in 
NGC~2023 and NGC~7023. 

\subsection{NGC~7023}

We carried out model fits to all three regions in the slit at the 
position NGC~7023 ISO-1 described in section 3. In the southern region, closest
to HD~200755, we only weakly detected the 1-0~S(0) and 1-0~S(1) lines and 
thus could not set useful limits on the physical parameters. At the 
position of the $\h2$ peak, and in the region north of the filament, we 
detected a number of lines. Our model fits to the $\h2-$peak constrain the 
density to be $n = 10^6\;\cm3$ for $G_0/n_H = 0.1$ and $0.01$. 
Draine \& Bertoldi's model fit to NGC 2023 (n2023b) 
with $n = 10^5\;\cm3$ for $G_0/n_H = 0.05$ was also a good fit to this region. 
The 1-0 S(1) line intensity is also roughly consistent with these 
physical parameters. 
For the north region, our model fits yield 
$n = 10^5\;\cm3$ for $G_0/n_H = 0.1$ and $n = 10^5 - 10^6\;\cm3$ for 
$G_0/n_H = 0.01$. The models with $n = 10^5\;\cm3$ for $G_0/n_H = 0.05$ 
were also a good fit to this region. The results for both of these regions 
agree with the density range and $G_0$ values ($n = 10^5 - 10^6\;\cm3$, $G_0 = 
10^3 - 10^4$) found by Martini et al. (1997), the value of $G_0 = 2.6 \times 
10^3$ found by Chokshi et al. (1988), and our line intensity measurement. 
The $\h2-$peak is at a projected distance of 0.1 pc from HD~200775, and so 
our results are also in agreement with the estimate that $G_0 = 300 - 1200$ 
at a distance of $0.1-0.2$ pc from HD~200775 by Federman et al. (1997). 

Our spectrum of NGC~7023 ISO-2 (not shown) showed only very weak
emission from the 1-0~S(1) line. We therefore did not have enough information 
to model the molecular hydrogen emission at this slit position. The strength 
of this emission, relative to that detected at NGC~7023 ISO-1 in a similar 
integration time, implies that the density or total column density of 
$\h2-$emitting material is less in this region. Also, given its greater
distance from the central star, in projection, this region is probably 
exposed to a lower $UV$ radiation field. 

The measured 1-0~S(1)/2-1~S(1) line ratios at the $\h2-$Peak and in the north 
region were 3.53 and 2.70, respectively. These two ratios are in broad 
agreement with the model fits described above. While the value at the 
$\h2-$Peak does imply a higher density than in the north region, an increase
in this ratio could also be due to changes in the gas temperature and $UV$ 
field strength. Both of these quantities could vary as the north region is 
farther, in projection, from the central star. The value of the 1-0~S(7)/H4 
ratio similarly decreases from the $\h2-$Peak to the north region, though 
changes in this ratio with density are also degenerate with changes in 
temperature and $UV$ field strength at densities approaching the critical 
density. 

Our measurements of $\trot$ and $\tvib$ at the $\h2-$Peak are in reasonable 
agreement with the measurement of Martini et al. (1997) at a different position
on the same filament. These results are higher than those found by 
Lemaire et al. (1996), but are within their uncertainties. 
Our measured $\trot = 860 \pm 230$ is also in agreement with the predictions 
of the best model fits to this position that we discuss above. We also find 
$\gamma$ to be consistent with both the observations of Martini et al. (1997)
and the models. 

\section{Continuum and $\h2$ Emission}

There are several models for the origin of the near-infrared continuum 
in reflection nebulae. One model suggests that the near-infrared continuum is 
due to thermal emission from tiny grains which are briefly heated by the 
absorption of single $UV$ photons (Sellgren, Werner, \& Dinerstein 1983; 
Sellgren 1984). Other models suggest a quasi-continuum of overlapping 
bands from polycyclic aromatic hydrocarbon (PAH) molecules (L\'eger \& 
Puget 1984; Allamandola, Tielens, \& Barker 1985), electronic fluorescence
in PAHs (Allamandola, Tielens, \& Barker 1989), or luminescence from 
hydrogenated amorphous carbon grains (Duley \& Williams 1988; Duley 1988). 

This continuum is not due to scattered starlight. 
Sellgren, Werner, \& Allamandola (1996), in a survey of 23 reflection nebulae, 
found that all of these objects have similar near-infrared colors, independent 
of the spectral type of the central star.  Sellgren, Werner, \& 
Dinerstein (1992) observed NGC~2023 and NGC~7023 in polarized light and 
calculated that less than 20\% of the $2.2\;\mu$m continuum emission, in 
contrast to the visible continuum, is due to scattered starlight. For 
NGC~7023, the small contribution of scattered light at $K$ ($4 - 5$\% 
polarization) compared to the larger scattered light contribution at $J$ 
($20 - 25$\% polarization) is confirmed by our spectra (Figures 8 \& 9), 
which show Pa$\beta$ emission at $1.28\;\mu$m scattered from the 
central Be star, but no Br$\gamma$ emission scattered from the Be star at
$2.17\;\mu$m. 

Our high signal-to-noise ratio, long-slit spectral observations of NGC~2023 and 
NGC~7023 are able to directly investigate the relative intensity of the 
$2.2\;\mu$m continuum and 1--0 S(1) line as shown in Figure 3.  These 
observations show that the $2.2\;\mu$m continuum peaks closer to the central 
star than the $\h2$ emission in both NGC~2023 and NGC~7023.  Our results at 
high spectral resolution confirm previous results from narrowband and 
broadband imaging of the near-infrared continuum and $\h2$ in NGC 2023 
(Gatley et al. 1987; Field et al. 1994) and NGC 7023 (Lemaire et al. 1996).
These images suggest that the near-infrared continuum peaks closer to the 
central star than does the fluorescent H$_2$ emission and that the 
near-infrared continuum emission is less filamentary than the H$_2$ emission.

The $3.3\;\mu$m emission feature, like the near-infrared continuum, has been 
proposed to arise from aromatic molecules or from aromatic grains, emitting 
either by fluorescence or by thermal emission during temperature fluctuations 
(see reviews by Puget \& L\'eger 1989; Allamandola et al. 1989; Sellgren 1994;
Papoular et al. 1996).  The observational relationship between the 
near-infrared continuum and the $3.3\;\mu$m feature, however, is unclear.
The near-infrared continuum and the $3.3\;\mu$m feature are always observed 
together in spectra of reflection nebula (Sellgren, Werner, \& Dinerstein 1983; 
Sellgren, Werner, \& Allamandola 1996).  Sellgren, Werner, \& Allamandola 
(1996) find that the ratio of the $3.3\;\mu$m feature to the nearby continuum 
is independent of the spectral type of the central star, but they also find 
approximately a factor of 3 variation in the feature-to-continuum ratio.
This variation in the feature-to-continuum ratio is found both within an 
individual source and among different sources.  Gatley et al. (1987) conclude 
from broadband imaging data that the near-infrared continuum in NGC~2023 is 
concentrated closer to the central star than is the $3.3\;\mu$m feature.
These observations suggest that while the materials and emission mechanisms 
giving rise to the near-infrared continuum and $3.3\;\mu$m emission feature may 
be similar, they are not identical.

Observations of the $3.3\;\mu$m emission feature in NGC~2023,
NGC~1333/SVS-3, and Parsamyan~18 show that the $3.3\;\mu$m feature is 
cospatial, in projection, with the $\h2$ emission in these three reflection 
nebulae (Gatley et al. 1987; Burton et al. 1989).  This is in contrast 
to results from regions with stronger UV fields, such as the Orion Nebula 
(Burton et al. 1989; Sellgren et al. 1990; Tielens et al. 1993) and the 
planetary nebula NGC~7027 (Graham et al. 1993), where the 3.3 $\mu$m emission
peaks closer to the UV source than the $\h2$ emission. In regions with 
stronger UV fields, the near-infrared continuum due to tiny grains
or large molecules is swamped by free-free emission from ionized hydrogen.
The absence of significant free-free emission in NGC~2023 and NGC~7023, 
however, allows a direct comparison of the location of the near-infrared 
continuum and $\h2$ emission.  In these two reflection nebulae the $2\;\mu$m 
continuum peaks closer to the star than the $\h2$ emission. 
Thus it appears that in reflection nebulae the material responsible for the 
near-infrared continuum is able to survive closer to the central star than the 
molecular gas. This places an interesting constraint on laboratory analogs 
and theoretical models for the continuum emission.

\section{Conclusions}

We have obtained near-infrared spectroscopy of four reflection nebulae, 
including NGC~1333 and NGC~2068 for which no previous near-infrared 
spectra have been analyzed. We have used these spectra to determine the 
physical environments of these four nebulae and derived their physical
properties by fitting the models of Draine \& Bertoldi (1996) to the observed 
line ratios. The three brightest nebulae in our sample, NGC~2023, NGC~2068, 
and NGC~7023, all show significant structure on scales of $\sim 2''$, 
corresponding to physical sizes of $\sim 1000$ AU at the distances of the 
nebulae. This structure corresponds to changes in density, continuum 
radiation, and the strength of the incident $UV$ field, and serves as 
direct evidence of the very inhomogeneous or clumpy nature of these PDRs. 
In addition to variations in the $\h2$ emission, we have spectroscopy of the 
underlying near-infrared continuum in NGC~2023 and NGC~7023 with sufficient 
signal-to-noise ratio to trace the peaks in surface brightness of the $\h2$
1-0~S(1) line relative to the continuum along our long slit. This analysis 
shows that the near-infrared continuum, particularly at $2.2\;\mu$m where 
scattered starlight is negligible, peaks closer to the central star than the 
$\h2$ emission in both nebulae. This result shows that the material 
responsible for the near-infrared continuum can survive closer to the central 
star than can the molecular hydrogen gas. 

\vskip 24pt

\acknowledgments

This work was supported in part by NASA grant NAG 5-3366 to KS. We would like 
to thank the staff of Lowell Observatory for their support during these 
observing runs and Rick Pogge for many helpful discussions.

\clearpage

\begin{center}
\begin{tabular}{ccc}
\multicolumn{3}{c}{{\bf TABLE 1}}\\[12pt]
\multicolumn{3}{c}{{\bf Summary of Observations}}\\[12pt]
\hline
\hline
\multicolumn{1}{c}{UT Date} &
\multicolumn{1}{c}{Source} & 
\multicolumn{1}{c}{Offset} \nl
\hline
1997 Sep 23 & NGC~7023 ISO-1	& $27''$W $34''$N \nl
1997 Sep 23 & NGC~1333 ISO-1	& $12''$E $26''$S \nl
1997 Oct 25 & NGC~1333 ISO-1	& \nl
1997 Oct 26 & NGC~7023 ISO-2	& $100''$N \nl
1997 Oct 26 & NGC~1333 ISO-1	& \nl
1997 Oct 27 & NGC~1333 ISO-1	& \nl
1997 Oct 28 & NGC~1333 ISO-1	& \nl
1998 Jan 30 & NGC~2068 C$-$Peak	& $72''$E \nl
1998 Jan 31 & NGC~2023 ISO-1 	& $60''$S \nl
1998 Feb 01 & NGC~2068 H$_{2}-$Peak & $121''$E \nl
1998 Feb 02 & NGC~2068 C$-$Peak	& \nl
\hline
\end{tabular}
\end{center}

\noindent
Table 1: The slit positions we observed for this study. Column 1 shows the 
date of the observation. In column 2 we list the name of the source, which is
a combination of the name of the nebula and a name describing the nebular 
position. Column 3 lists the offset from the central star of the nebula. This
offset position is the center of the $4.5'' \times 45''$ slit oriented
North-South. See Section 3 for further details.  

\clearpage

\begin{center}
\begin{tabular}{clc}
\multicolumn{3}{c}{{\bf TABLE 2}}\\[12pt]
\multicolumn{3}{c}{{\bf J and H Band Line Blends}}\\[12pt]
\hline
\hline
\multicolumn{1}{c}{Blend} &
\multicolumn{1}{c}{Line} &
\multicolumn{1}{c}{Wavelength} \nl
\hline
J1 (1.307 - 1.322 $\mu$m) & 5-3 S(5)  & 1.31067 \nl
   & 4-2 S(1)  & 1.31157 \nl
   & 3-1 Q(1)  & 1.31410 \nl
   & 9-6 Q(1)  & 1.31582 \nl
   & 3-1 Q(2)  & 1.31807 \nl
   & 2-0 Q(9)  & 1.31877 \nl
   &           &         \nl
H1 (1.556 - 1.566 $\mu$m) & 5-3 O(2)  & 1.56073 \nl
   & 7-5 S(3)  & 1.56150 \nl
   & 5-3 Q(7)  & 1.56263 \nl
   & 4-2 O(4)  & 1.56352 \nl
   &           &         \nl
H2 (1.598 - 1.624 $\mu$m) & 6-4 Q(1)  & 1.60153 \nl
   & 6-4 Q(2)  & 1.60739 \nl 
   & 5-3 Q(9)  & 1.60839 \nl
   & 4-2 Q(13) & 1.61225 \nl
   & 5-3 O(3)  & 1.61354 \nl
   & 6-4 Q(3)  & 1.61621 \nl
   & 7-5 S(1)  & 1.62053 \nl
   & 4-2 O(5)  & 1.62229 \nl
   &           &         \nl
H3 (1.666 - 1.678 $\mu$m) & 6-4 O(2) & 1.67502 \nl
   & 5-3 O(4)  & 1.67182 \nl
   & 6-4 O(2)  & 1.67502 \nl
   &           &         \nl
H4 (1.724 - 1.740 $\mu$m) & 7-5 Q(1)  & 1.72878 \nl
   & 8-6 S(2)  & 1.72967 \nl
   & 6-4 O(3)  & 1.73264 \nl
   & 7-5 Q(2)  & 1.73573 \nl
   & 5-3 O(5)  & 1.73589 \nl
   & 6-4 Q(9)  & 1.73695 \nl
\hline
\end{tabular}
\end{center}

\noindent
Table 2: The prominent $\h2$ line blends in the $J-$ and $H-$bands in these 
spectra along with their principal components. 
Column 1 lists the designation we have adopted for each blend along
with the wavelength range we integrated over to measure the blend intensity.
Columns 2 and 3 list the transitions and wavelengths of all lines 
brighter than 1\% of 1-0 S(1).

\clearpage

\begin{center}
\begin{tabular}{cccccccc}
\multicolumn{8}{c}{{\bf TABLE 3}}\\[12pt]
\multicolumn{8}{c}{{\bf NGC~1333 and NGC~2023 Line Intensities}}\\[12pt]
\hline
\hline
\multicolumn{1}{c}{Line ID} &
\multicolumn{1}{c}{Wavelength} & 
\multicolumn{1}{c}{${\rm I_{NGC\;1333}}$} & 
\multicolumn{1}{c}{$\sigma_I$} &
\multicolumn{1}{c}{${\rm I_{NGC\;2023\;South}}$} & 
\multicolumn{1}{c}{$\sigma_I$} &
\multicolumn{1}{c}{${\rm I_{NGC\;2023\;North}}$} & 
\multicolumn{1}{c}{$\sigma_I$} \nl
\hline
3-2 S(2) & 2.28703 & 0.142 & 0.078 &       &       & 0.080 & 0.036 \nl
2-1 S(1) & 2.24772 & 0.451 & 0.060 & 0.401 & 0.067 & 0.368 & 0.043 \nl
1-0 S(0) & 2.22330 & 0.703 & 0.065 & 0.465 & 0.062 & 0.472 & 0.044 \nl
3-2 S(3) & 2.20140 & 0.163 & 0.057 & 0.155 & 0.059 & 0.162 & 0.029 \nl
2-1 S(2) & 2.15423 & 0.183 & 0.058 & 0.155 & 0.059 & 0.156 & 0.033 \nl
1-0 S(1) & 2.12183 & 1.000 & 0.045 & 1.000 & 0.043 & 1.000 & 0.028 \nl
2-1 S(3) & 2.07351 & 0.248 & 0.046 & 0.315 & 0.097 & 0.293 & 0.030 \nl
1-0 S(2) & 2.03376 & 0.419 & 0.060 & 0.422 & 0.097 & 0.398 & 0.067 \nl
1-0 S(7) & 1.74803 & 0.122 & 0.069 & 0.237 & 0.083 & 0.201 & 0.073 \nl
6-4 Q(1) & 1.60153 & 0.130 & 0.073 & 0.107 & 0.064 & 0.104 & 0.055 \nl
4-2 S(3) & 1.26155 & 0.268 & 0.135 & 0.290 & 0.121 & 0.244 & 0.092 \nl
J1	 & 	   & 0.698 & 0.218 & 0.724 & 0.262 & & \nl
H1	 & 	   & 0.324 & 0.095 & 0.212 & 0.059 & 0.177 & 0.071 \nl
H2	 & 	   & 0.675 & 0.214 & 0.480 & 0.152 & 0.580 & 0.215 \nl
H3	 & 	   & 0.263 & 0.082 & & & & \nl
H4	 & 	   & 0.509 & 0.116 & 0.390 & 0.106 & 0.303 & 0.119 \nl
\hline
\end{tabular}
\end{center}

\noindent
Table 3: Measured line intensities for NGC~1333 and NGC~2023 relative to 
the 1-0~S(1) line. Column~1 lists the identification for each line, while 
column 2 lists the central wavelength of the feature. Columns 3 \& 4 contain
the intensity and $1\sigma$ uncertainty for NGC~1333. Columns 5, 6, 7, \& 8 
contain the corresponding information for the south and north regions of 
NGC~2023, respectively. These intensity measurements have not been 
corrected for extinction by dust. The $1\sigma$ uncertainties for the 1-0~S(7), 
6-4~Q(1), and 4-3~S(3) lines and all of the line blends also contain an 
additional 5\% error to account for the uncertainty in the scaling from $J$ 
and $H$ to the $K-$band. 

\clearpage

\begin{center}
\begin{tabular}{cccccccccc}
\multicolumn{6}{c}{{\bf TABLE 4}}\\[12pt]
\multicolumn{6}{c}{{\bf NGC~2068 Line Intensities}}\\[12pt]
\hline
\hline
\multicolumn{1}{c}{Line ID} &
\multicolumn{1}{c}{Wavelength} &
\multicolumn{1}{c}{${\rm I_{NGC\;2068-C}}$} &
\multicolumn{1}{c}{$\sigma_I$} &
\multicolumn{1}{c}{${\rm I_{NGC\;2068-\h2}}$} &
\multicolumn{1}{c}{$\sigma_I$} \nl
\hline
2-1 S(1) & 2.24772 & 0.496 & 0.195 & 0.376 & 0.066 \nl
1-0 S(0) & 2.22330 & 0.806 & 0.202 & 0.488 & 0.067 \nl
2-1 S(2) & 2.15423 & 0.374 & 0.180 & 0.183 & 0.051 \nl
1-0 S(1) & 2.12183 & 1.000 & 0.149 & 1.000 & 0.052 \nl
2-1 S(3) & 2.07351 & 0.311 & 0.151 & 0.275 & 0.093 \nl
1-0 S(2) & 2.03376 & 0.554 & 0.231 & 0.466 & 0.095 \nl
1-0 S(7) & 1.74803 &       &       & 0.182 & 0.080 \nl
6-4 Q(1) & 1.60153 &       &       & 0.132 & 0.072 \nl
J1	 & 	   &	   &	   & 0.363 & 0.143 \nl 
H1	 &	   &	   &	   & 0.219 & 0.064 \nl
H2	 &	   & 0.761 & 0.370 & 0.711 & 0.195 \nl
H3	 &	   & 0.239 & 0.114 &       &       \nl
H4	 &	   & 0.797 & 0.247 & 0.396 & 0.950 \nl
\hline
\end{tabular}
\end{center}

\noindent
Table 4: As in Table 3, but for NGC~2068.

\begin{center}
\begin{tabular}{cccccc}
\multicolumn{6}{c}{{\bf TABLE 5}}\\[12pt]
\multicolumn{6}{c}{{\bf NGC~7023 Line Intensities}}\\[12pt]
\hline
\hline
\multicolumn{1}{c}{Line ID} &
\multicolumn{1}{c}{Wavelength} &
\multicolumn{1}{c}{${\rm I_{NGC\;7023\;Peak}}$} &
\multicolumn{1}{c}{$\sigma_I$} &
\multicolumn{1}{c}{${\rm I_{NGC\;7023\;North}}$} &
\multicolumn{1}{c}{$\sigma_I$} \nl
\hline
2-1 S(1) & 2.24772 & 0.290 & 0.091 & 0.380 & 0.161 \nl
1-0 S(0) & 2.22330 & 0.411 & 0.088 & 0.516 & 0.164 \nl
3-2 S(3) & 2.20140 & 0.150 & 0.049 & & \nl
2-1 S(2) & 2.15423 & 0.138 & 0.061 & & \nl
1-0 S(1) & 2.12183 & 1.000 & 0.060 & 1.000 & 0.087 \nl
2-1 S(3) & 2.07351 & 0.182 & 0.038 & 0.241 & 0.111 \nl
1-0 S(2) & 2.03376 & 0.335 & 0.052 & 0.411 & 0.150 \nl
1-0 S(7) & 1.74803 & 0.213 & 0.057 & 0.471 & 0.126 \nl
H1	 & 	   & 0.140 & 0.067 & 0.372 & 0.132 \nl
H2	 & 	   & 0.347 & 0.067 & 0.958 & 0.364 \nl
H4	 & 	   & 0.209 & 0.100 & 0.637 & 0.301 \nl
\hline
\end{tabular}
\end{center}

\noindent
Table 5: As in Table 3, but for NGC~7023.

\clearpage

\begin{center}
\begin{tabular}{lcccccc}
\multicolumn{7}{c}{{\bf TABLE 6}}\\[12pt]
\multicolumn{7}{c}{{\bf PDR Diagnostics}}\\[12pt]
\hline
\hline
\multicolumn{1}{c}{Nebula} &
\multicolumn{1}{c}{$\frac{1-0S(1)}{2-1S(1)}$} &
\multicolumn{1}{c}{$\frac{2-1S(1)}{6-4Q(1)}$} &
\multicolumn{1}{c}{$\frac{H4}{1-0S(7)}$} &
\multicolumn{1}{c}{$\trot$} &
\multicolumn{1}{c}{$\tvib$} &
\multicolumn{1}{c}{$\gamma$} \nl
\hline
NGC~1333            & 2.26 & 3.02 & 4.20 &  690 & 5000 & 1.4 \nl
	      & $\pm$ 0.32 & 1.57 & 2.38 &  120 &  720 & 0.4 \nl
NGC~2023 North      & 2.53 & 3.37 & 1.65 &  920 & 4540 & 1.8 \nl
	      & $\pm$ 0.44 & 1.93 & 0.69 &  250 &  800 & 0.7 \nl
NGC~2023 South      & 2.76 & 3.16 & 1.52 &  870 & 4240 & 1.9 \nl
              & $\pm$ 0.34 & 1.57 & 0.76 &  170 &  520 & 0.5 \nl
NGC~2068 C-Peak     & 2.11 &      &      &      &      &     \nl
              & $\pm$ 0.91 &      &      &      &      &     \nl
NGC~2068 $\h2-$Peak & 2.78 & 2.07 & 2.20 & 1000 & 4220 & 1.7 \nl
              & $\pm$ 0.53 & 0.94 & 0.92 &  240 &  800 & 0.6 \nl
NGC~7023 $\h2-$Peak & 3.53 &      & 0.98 &  860 & 3570 & 2.2 \nl
 	      & $\pm$ 1.15 &      & 0.49 &  230 & 1170 & 0.8 \nl
NGC~7023 North      & 2.70 &      & 1.36 &      &      &     \nl
              & $\pm$ 1.19 &      & 0.67 &      &      &     \nl
\hline
\end{tabular}
\end{center}

\noindent
Table 6: Line ratios, temperatures, and ortho-to-para ratios for the 
nebulae discussed in this paper. Column 1 lists the nebula (including the 
offset position), while columns 2, 3 \& 4 list the line ratios of the 
1-0~S(1) to 2-1~S(1) lines, 2-1~S(1) to 6-4~Q(1) lines, and H4 blend to 
1-0~S(7) line, respectively. The rotation temperature for $v = 1, J = 2, 4$ 
and the vibration temperature for $v = 1, 2, J = 3$ are listed in columns 
5 \& 6 in units of Kelvin, while column 7 contains the ratio of ortho$-$ to 
para$-\h2$ based on the relative column densities in the $v = 1, J = 2 - 4$ 
levels. The $1\sigma$ uncertainties in all of these quantities are 
tabulated immediately below them.  All of the values listed in this Table 
have been corrected for interstellar reddening as described in section 4. 

\clearpage

\begin{figure}
\plotfiddle{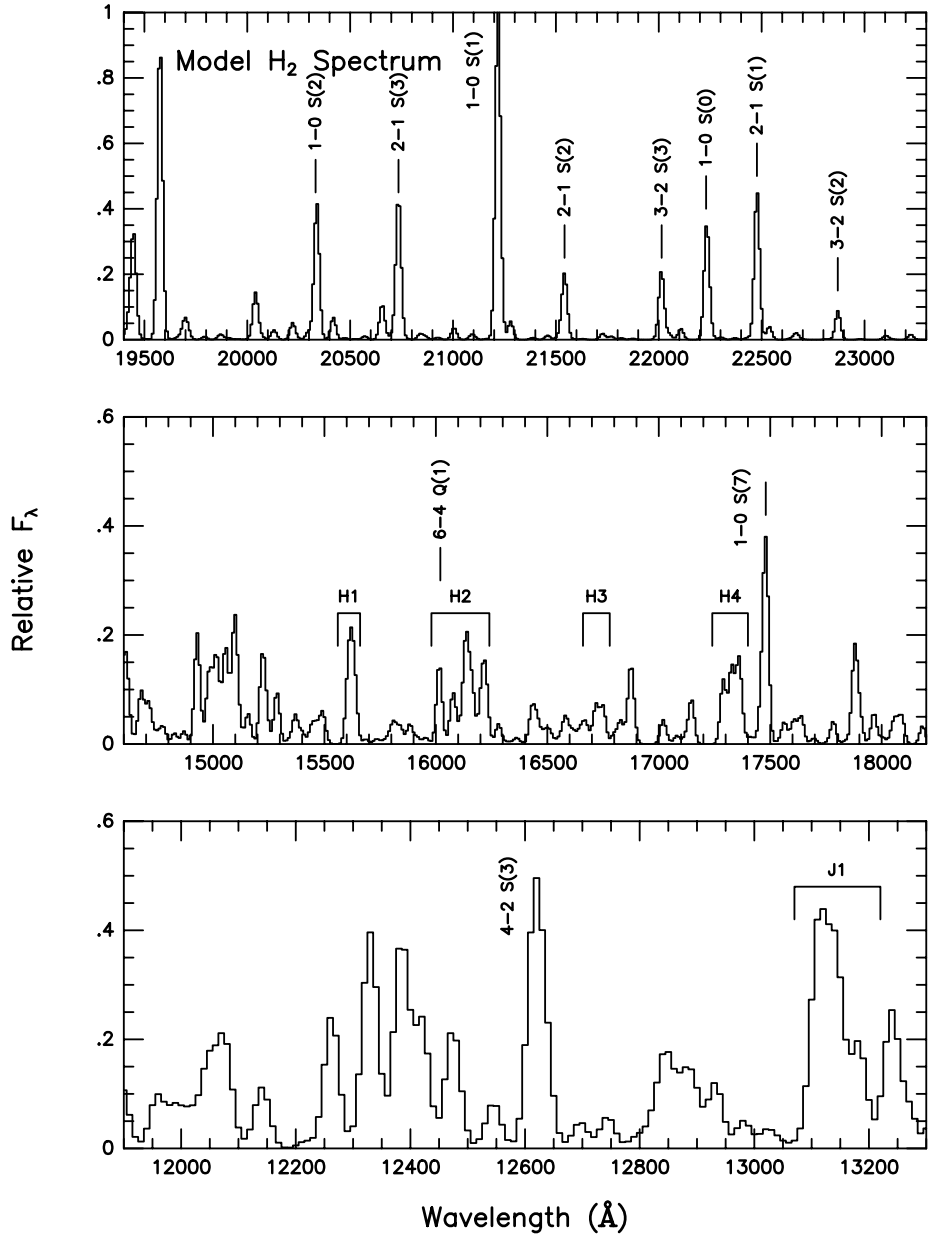}{5.0truein}{0}{70}{70}{-220}{-40}
\caption{A model spectrum of $\h2$ emission (Draine \& Bertoldi 1996) for a 
density of $10^5\;\cm3$ and incident $UV$ field $10^4$ times the ambient 
interstellar field that has been smoothed to our instrumental resolution. 
We show the near-infrared $J-$, $H-$, and $K-$bands over the same spectral 
range as our data and have marked the prominent lines and blends detected in 
most of the objects we observed. }
\end{figure}

\clearpage

\begin{figure}
\plotfiddle{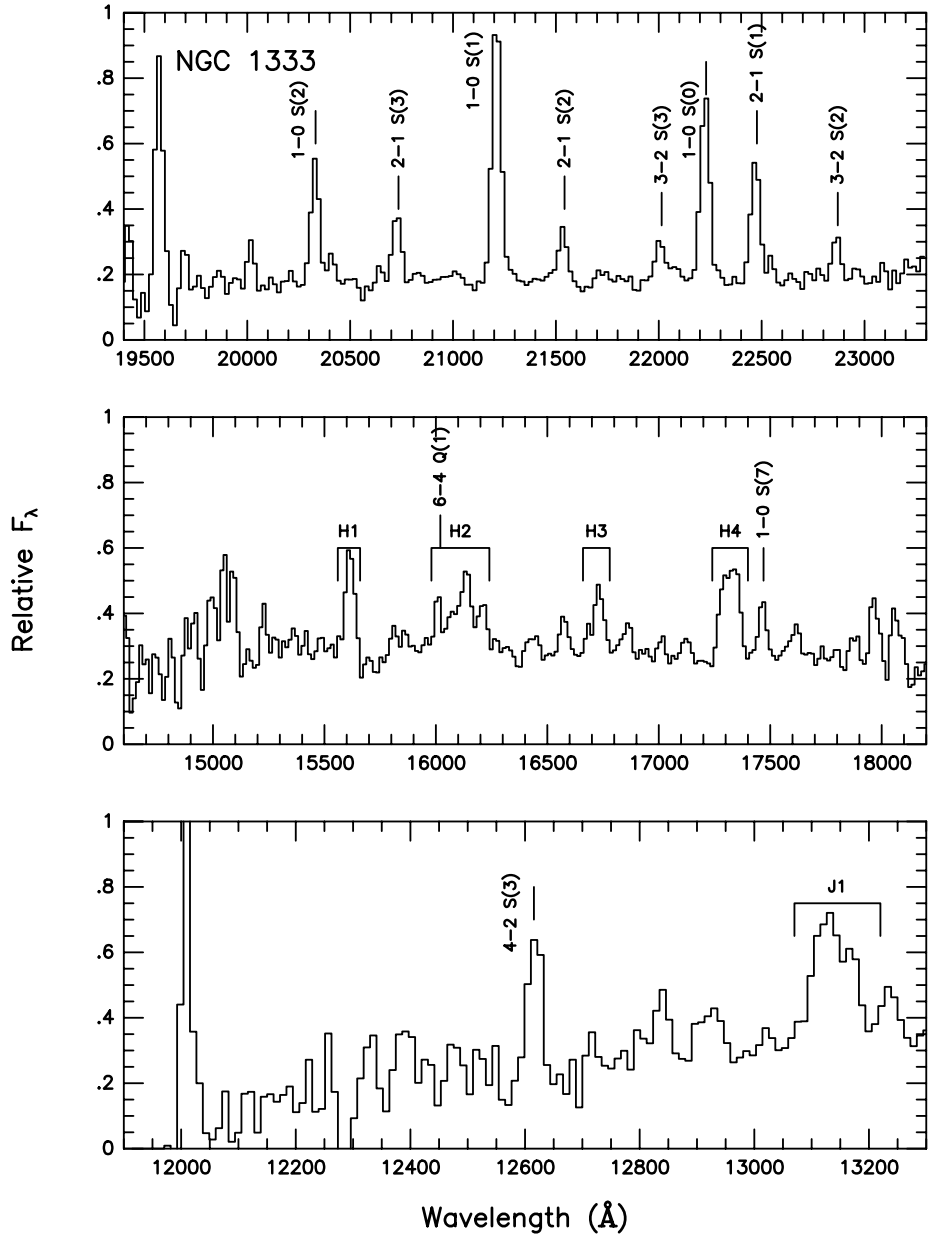}{5.0truein}{0}{70}{70}{-220}{-40}
\caption{Spectrum of NGC~1333 in the near-infrared $J-$, $H-$, and $K-$bands. 
This spectrum was obtained at a position $12''$ east, $26''$ south of 
BD~$+30\ 549$ and is a summation of all $45''$ of the $4.5''-$wide, north-south
slit (see section 3.1). The spectra are all on the same relative 
F$_{\lambda}$ (flux per unit wavelength) scale as described in section 2. }
\end{figure}

\clearpage

\begin{figure}
\plotfiddle{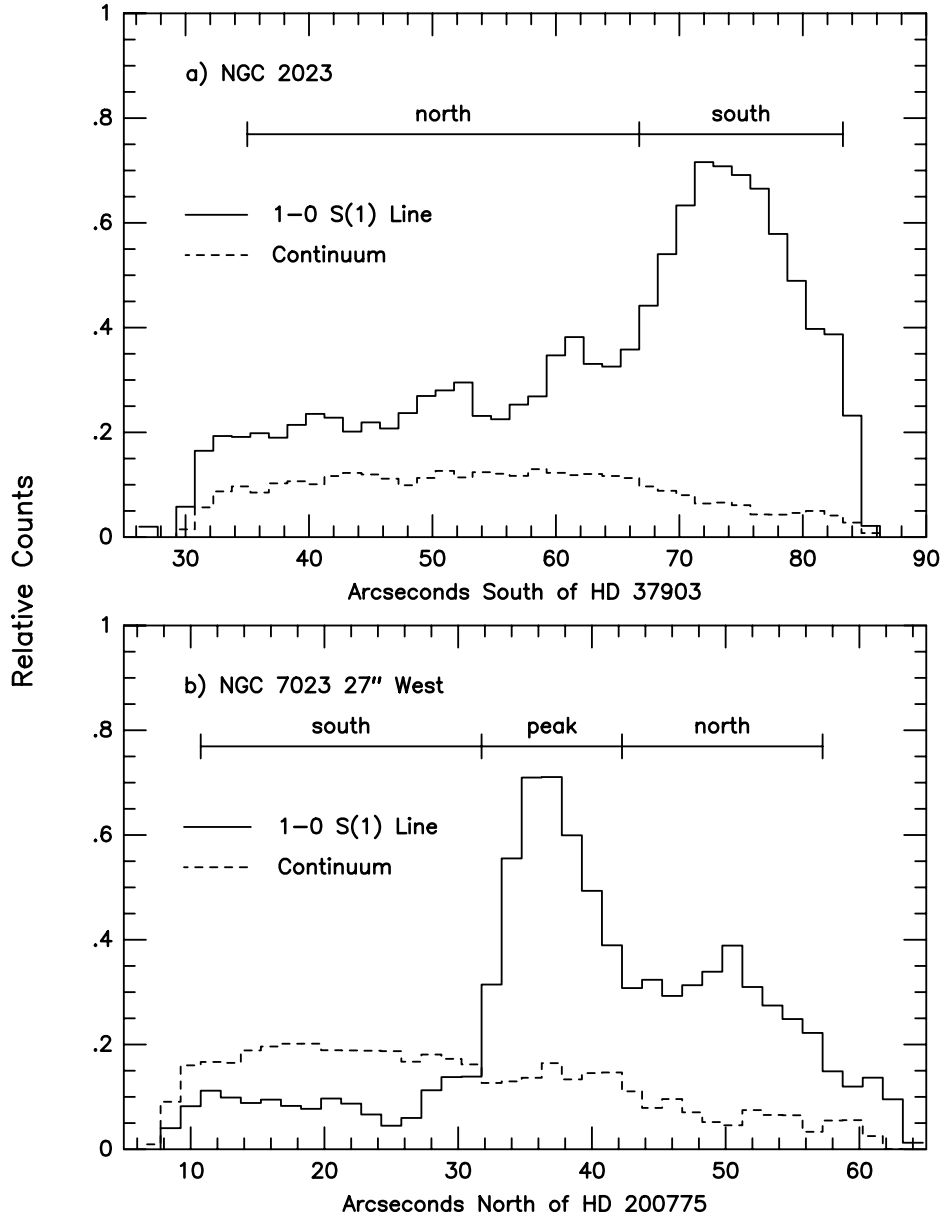}{5.0truein}{0}{70}{70}{-220}{-60}
\caption{Continuum and 1-0 S(1) line intensity along the $4.5''$ wide,
north-south slit for a) NGC~2023 and b) NGC~7023. The 1-0 S(1) counts were 
obtained from a 60 \AA\ wide bin centered on the 1-0 S(1) line at 2.121 $\mu$m. 
The continuum counts are the average of two 60 \AA\ wide bins on either side of 
the 1-0 S(1) emission line. These counts have not been corrected for the slit 
illumination function. a) Our slit was centered $60''$ south of HD~37903 with 
no east-west offset. The two regions marked in the figure are described in 
more detail in section 3.2. b) For NGC~7023 the slit was centered $27''$ west, 
$34''$ north of HD~200775 and the three regions are described in more detail 
in section 3.4. }
\end{figure}

\clearpage

\begin{figure}
\plotfiddle{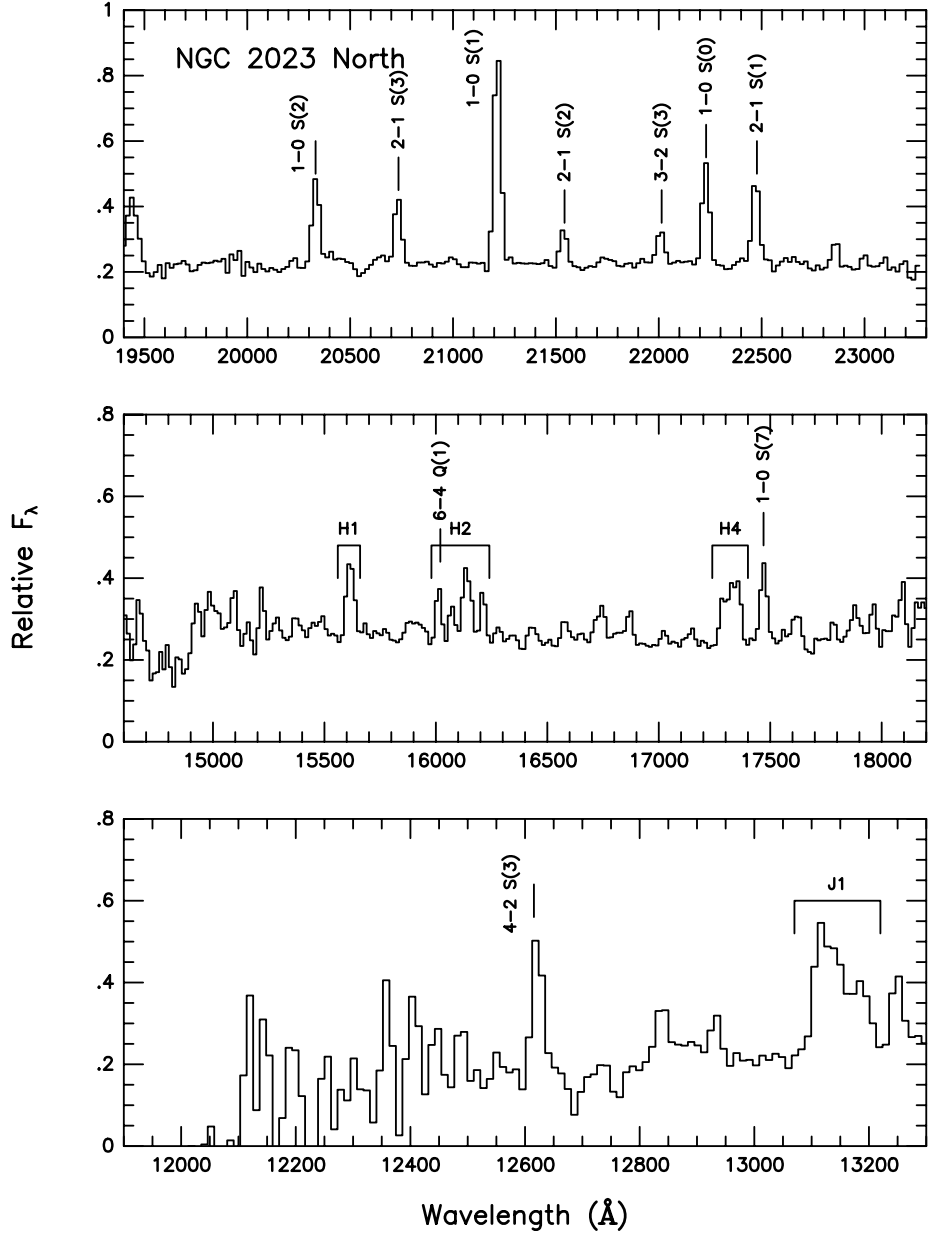}{5.0truein}{0}{70}{70}{-220}{-40}
\caption{As in Figure 2 for the `north region' of NGC~2023 (see section 3.2). 
This region extends over $35'' - 67''$ south, $0''$ east of HD~37903 and is 
bright in continuum emission (see also Figure 3).}
\end{figure}

\clearpage

\begin{figure}
\plotfiddle{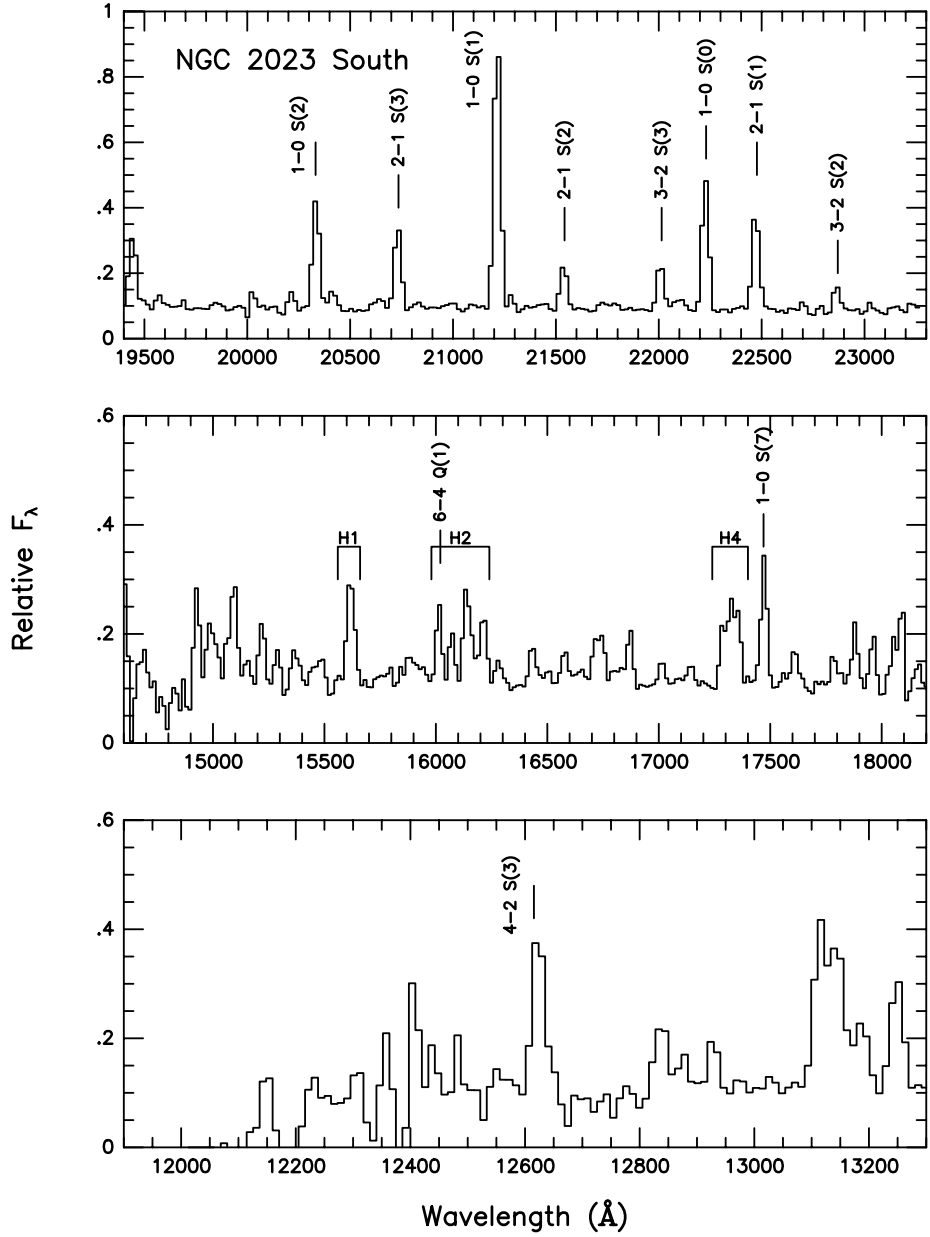}{5.0truein}{0}{70}{70}{-220}{-40}
\caption{As in Figure 2 for the 'south region' of NGC~2023 (see section 3.2). 
This region extends over $68'' - 84''$ south, $0''$ east 
and exhibits stronger $\h2$ emission than the north region shown in Figure 4, 
but weaker continuum emission (see also Figure 3). }
\end{figure}

\clearpage

\begin{figure}
\plotfiddle{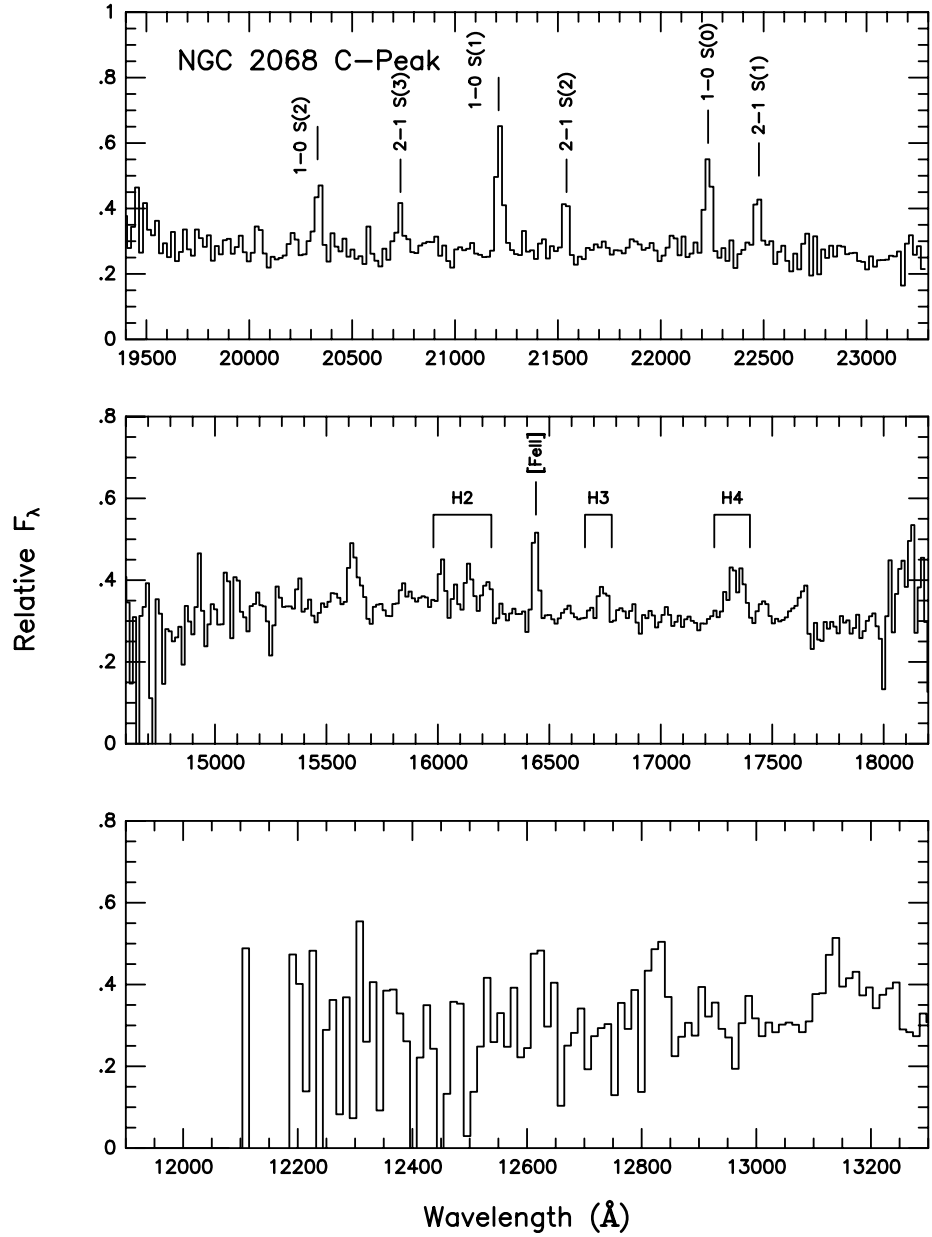}{5.0truein}{0}{70}{70}{-220}{-40}
\caption{As in Figure 2 for the continuum peak in NGC 2068, which is 
centered 72$''$ due east of HD~38563-N (see section 3.3). This spectrum is the 
summation of the entire 45$''$ of the slit. }
\end{figure}

\clearpage

\begin{figure}
\plotfiddle{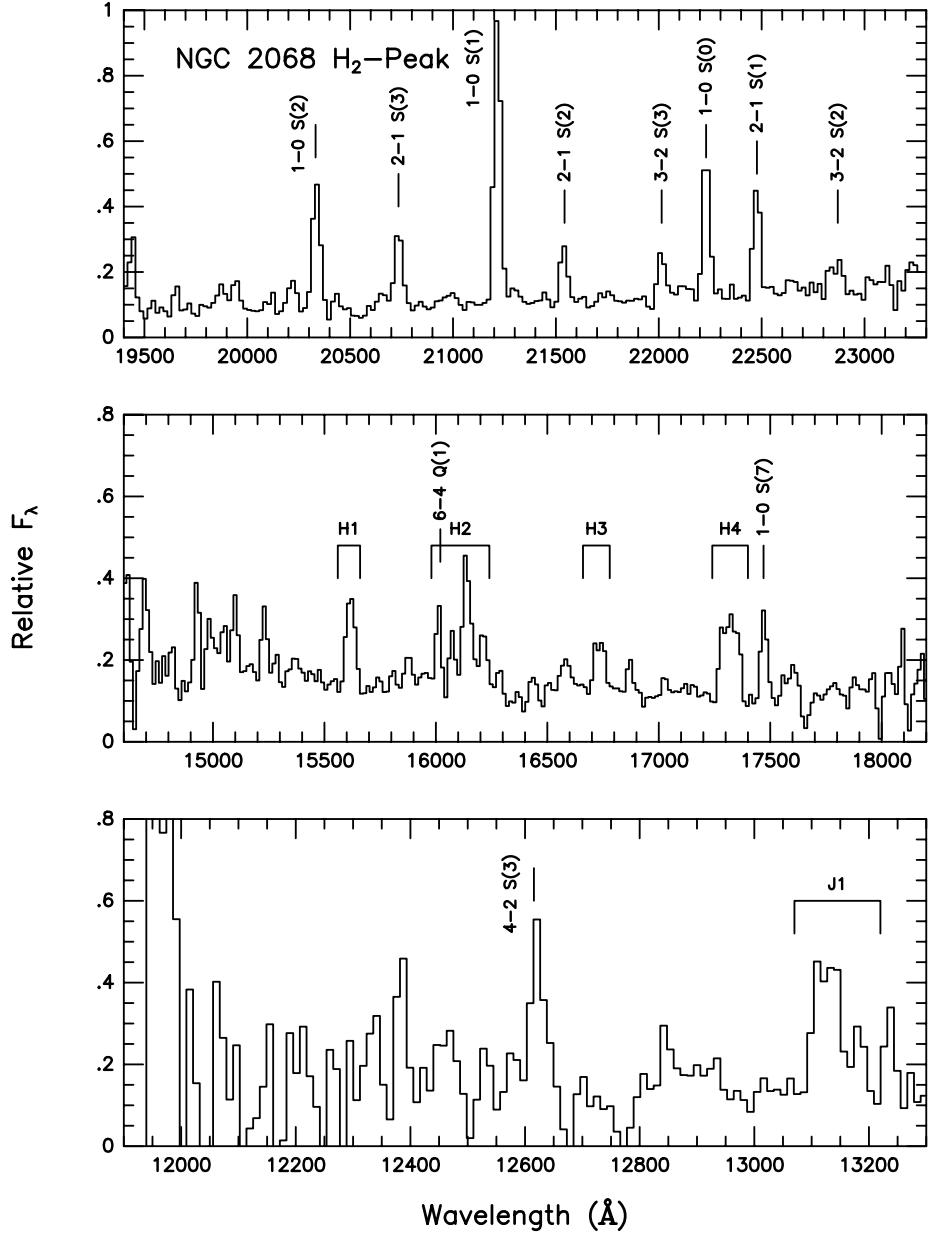}{5.0truein}{0}{70}{70}{-220}{-40}
\caption{As in Figure 2 for the $\h2-$Peak in NGC~2068, which is centered 
121$''$ due east of HD~38563-N (see section 3.3). This spectrum is the sum of
the slit over $0'' - 23''$ north of the central double star. }
\end{figure}

\clearpage

\begin{figure}
\plotfiddle{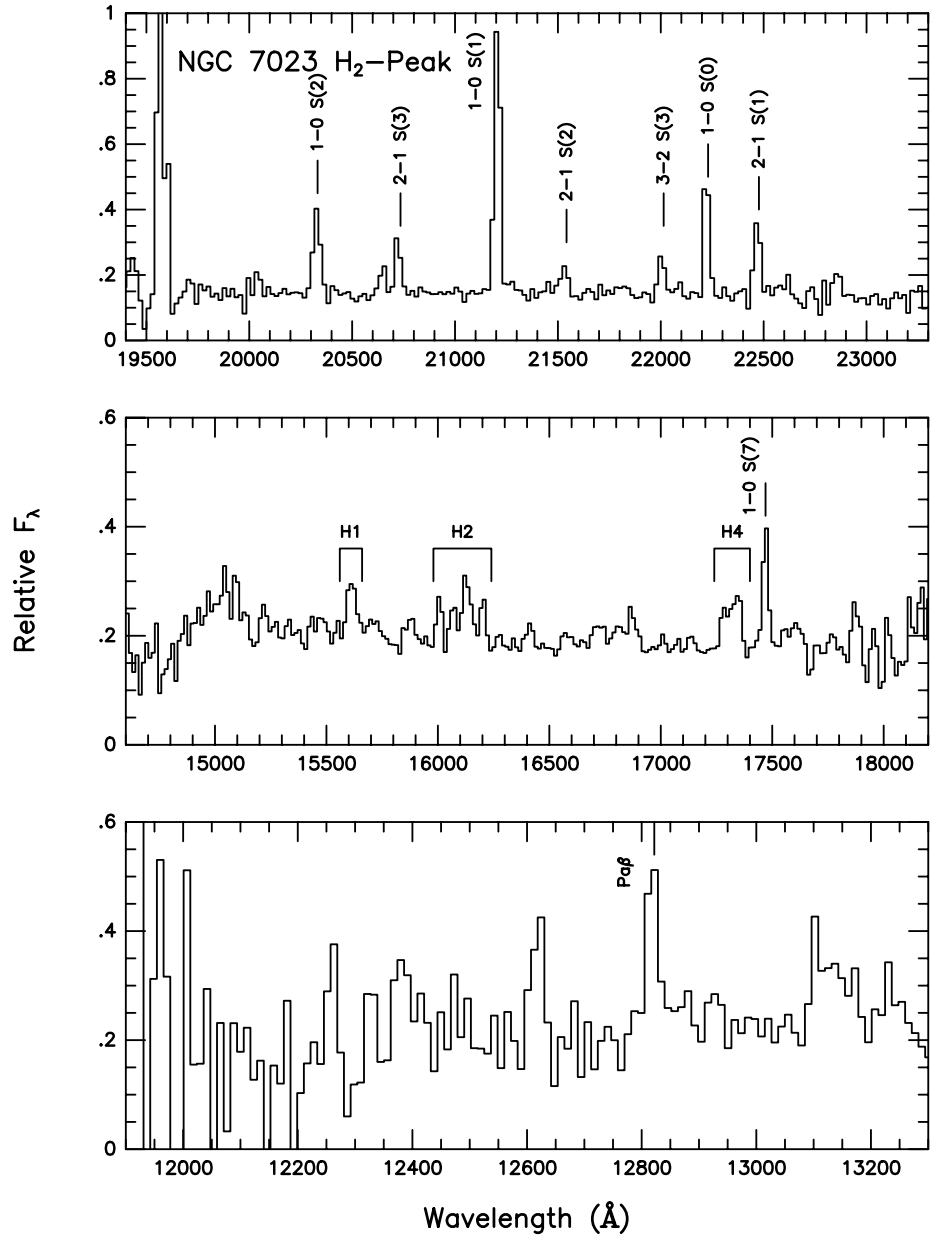}{5.0truein}{0}{70}{70}{-220}{-40}
\caption{As in Figure 2 for the $\h2-$Peak in NGC~7023 (section 3.4). 
This region extends over $32'' - 41''$ north at an offset $34''$ west of 
HD~200775 and contains the strongest $\h2$ and continuum emission. }
\end{figure}

\clearpage

\begin{figure}
\plotfiddle{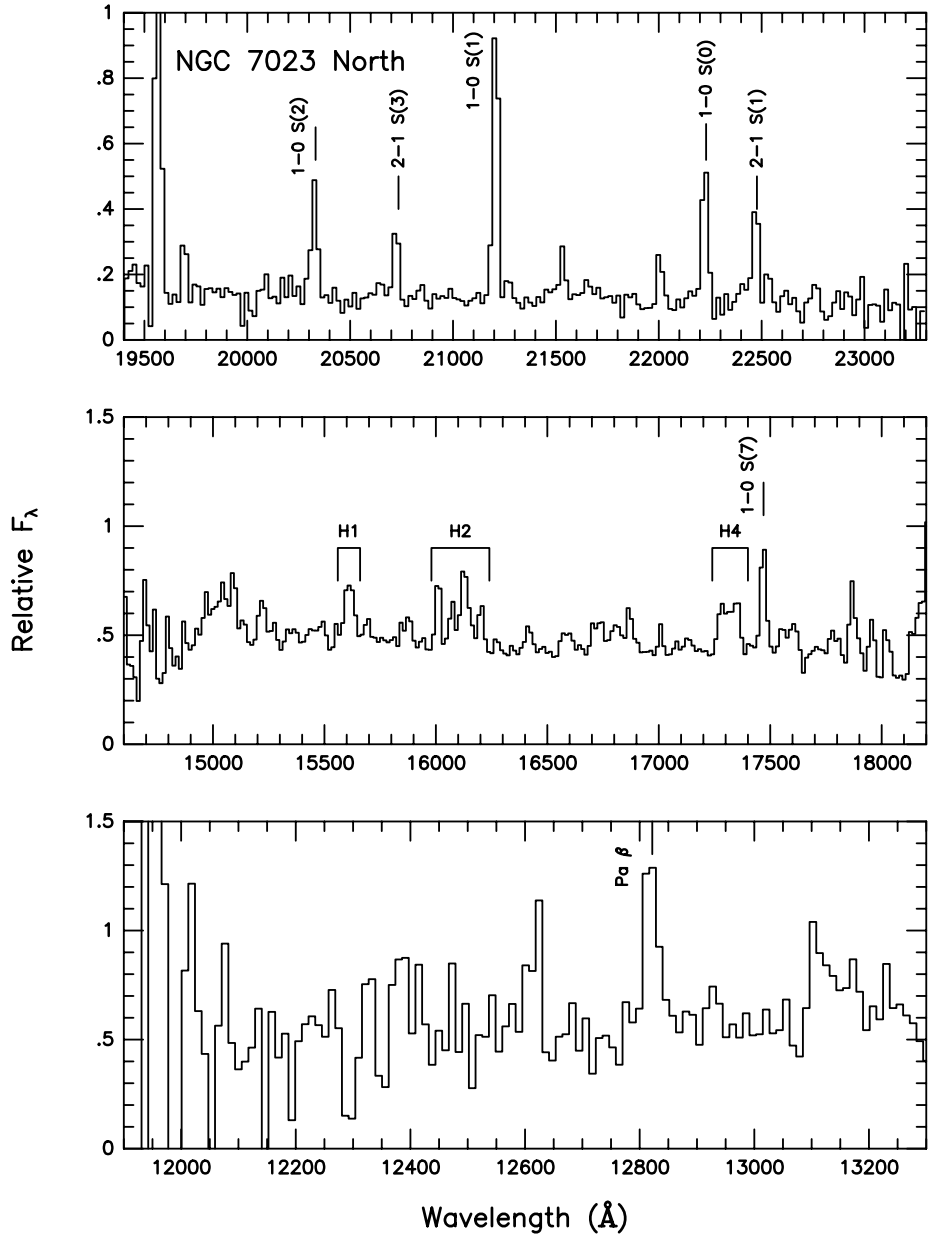}{5.0truein}{0}{70}{70}{-220}{-40}
\caption{As in Figure 2 for the `north region' in NGC~7023 (see section 3.4). 
This region extends over $41'' - 56''$ north at an offset $34''$ west of 
HD~200775. This region still exhibits bright $\h2$ emission, but the continuum 
intensity is significantly less than the $\h2-$Peak (Figure 8) to the south 
(see also Figure 2). }
\end{figure}
\clearpage

\begin{figure}
\plotfiddle{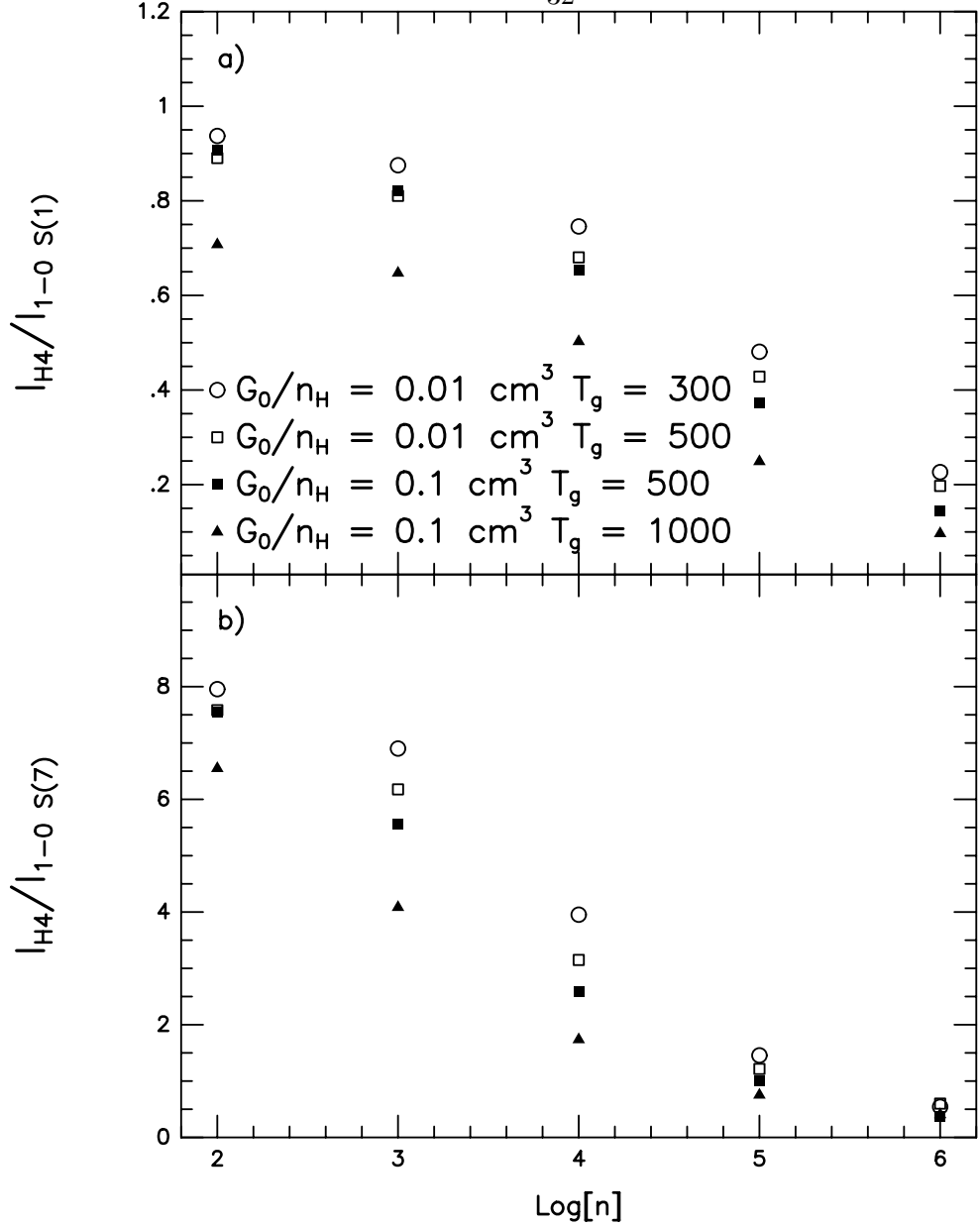}{5.0truein}{0}{70}{70}{-220}{-50}
\caption{In a) we show how the intensity of the H4 blend (see Table 2), 
normalized to the 1-0 S(1) line intensity, varies as a function of the 
PDR parameters in the models of Draine \& Bertoldi (1996). The behavior of
this blend is representative of the other $\h2$ blends listed in Table 2 as 
all of the blends are dominated by transitions from high vibrational levels. 
In b) we plot the ratio of the H4 blend to the 1-0 S(7) line. This pair of 
features is a relatively good diagnostic of the PDR and has the advantage of 
being insensitive to the amount of reddening. Our observed values of this ratio 
vary from 0.98 to 4.20 and fall within the range predicted by the models.}
\end{figure}
\clearpage

\begin{figure}
\plotfiddle{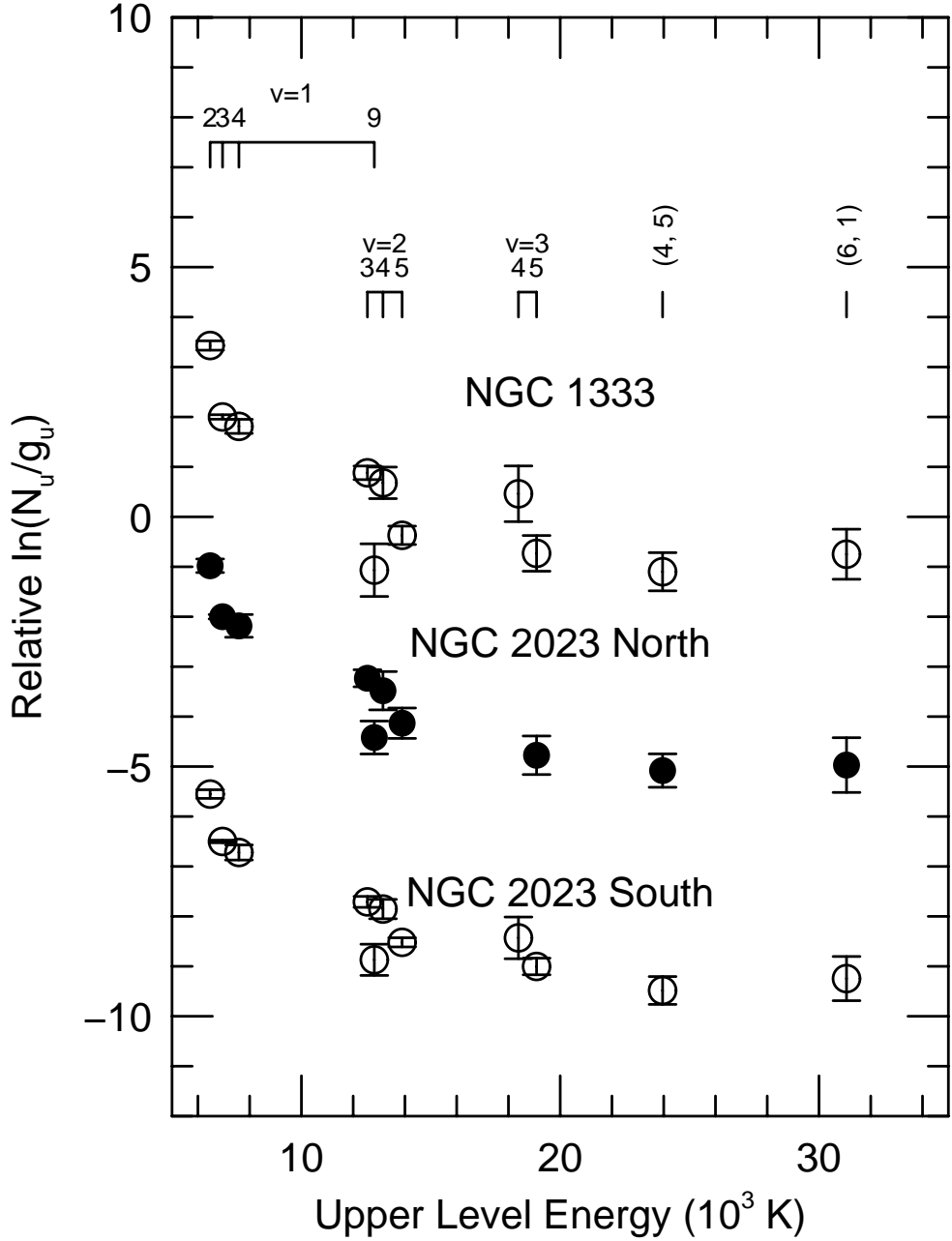}{5.0truein}{0}{70}{70}{-220}{-40}
\caption{Relative Population Diagrams for NGC~1333 and the two slit 
regions in NGC~2023 plotted as a function of the upper level energy divided 
by the Boltzman constant. The labels at the top of the Figure provide the
rotational ($J$) and vibrational ($v$) quantum number of the level giving rise 
to the observed transition. The quantum numbers in parenthesis, e.g. (4,5), 
are ($v, J$). N$_u$ is the column density of the upper level, while g$_u$ 
is its statistical weight (see section 4).  }
\end{figure}

\clearpage

\begin{figure}
\plotfiddle{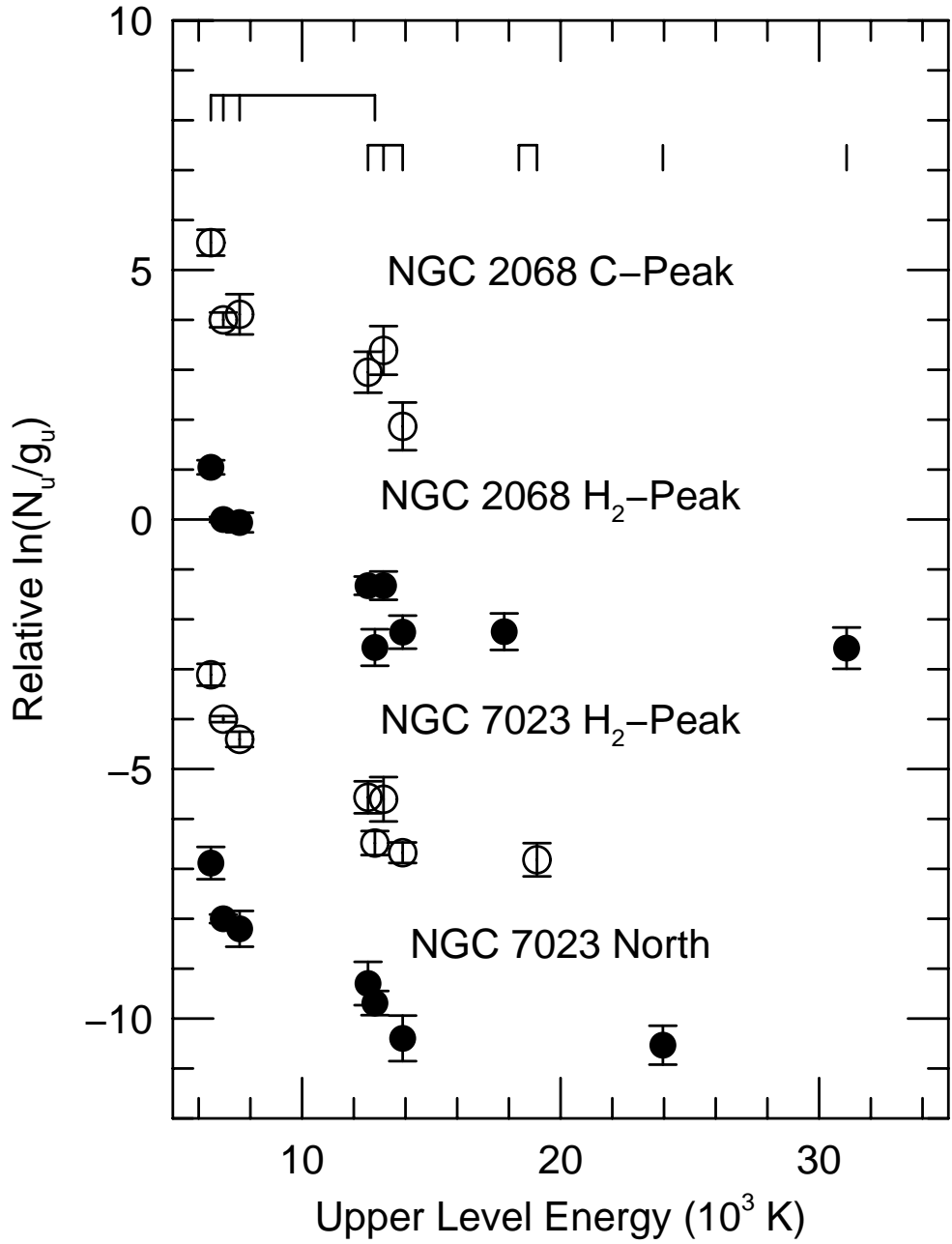}{5.0truein}{0}{70}{70}{-220}{-40}
\caption{Same as Figure 11 but for the continuum and $\h2-$Peak in NGC~2068 
and the $\h2-$Peak and north region in NGC~7023. }
\end{figure}

\end{document}